\pdfoutput=1
\RequirePackage{amsmath}
\documentclass[traditabstract]{aa}
\usepackage{natbib,amssymb,txfonts,astro,url,mathtools}
\usepackage[version=3]{mhchem}
\usepackage[normalem]{ulem}
\usepackage{longtable,booktabs}
 \usepackage{lscape}
\usepackage{diagbox}
\usepackage{cellspace}
\usepackage{slashbox}
\usepackage{multirow}

\newcolumntype{C}{Sc}
\usepackage{color}
\DeclareGraphicsExtensions{.pdf,.png,.eps,.gif,.ps}
\begin{document}
\setlength{\mathindent}{0pt}
\titlerunning {The ortho-to-para ratio of interstellar NH$_2$}
\title{The ortho-to-para ratio of interstellar NH$_2$: Quasi-classical
  trajectory calculations and new simulations} \authorrunning {R.~Le Gal et al}

\author{R.~Le~Gal\inst{1}, E.~Herbst\inst{1}, C. Xie\inst{2},
  A. Li\inst{3}, H.~Guo\inst{2}}

\institute{$^1$ Department of Chemistry, University of Virginia,
  McCormick Road, Charlottesville, VA 22904, USA,
  \email{\url{romane.legal@virginia.edu}} \\ $^2$ Department of
  Chemistry and Chemical Biology, University of New Mexico,
  Albuquerque, New Mexico, 87131, USA\\ $^3$ Key Laboratory of
  Synthetic and Natural Functional Molecule Chemistry, Ministry of
  Education, College of Chemistry and Materials Science, Northwest
  University, Xi’an 710127, China }

 \date{Received 13 June 2016; Accepted 6 September 2016}

 \abstract{ Based on recent \emph{Herschel} results, the ortho-to-para
   ratio (OPR) of NH$_2$ has been measured towards the following
   high-mass star-forming regions: W31C (G10.6-0.4), W49N (G43.2-0.1),
   W51 (G49.5-0.4), and G34.3+0.1. The OPR at thermal equilibrium
   ranges from the statistical limit of three at high temperatures to
   infinity as the temperature tends toward zero, unlike the case of
   H$_{2}$. Depending on the position observed along the
   lines-of-sight, the OPR was found to lie either slightly below the
   high temperature limit of three (in the range 2.2-2.9) or above
   this limit ($\sim3.5$, $\gtrsim 4.2$, and $\gtrsim 5.0$). In low
   temperature interstellar gas, where the H$_{2}$ is para-enriched,
   our nearly pure gas-phase astrochemical models with
   nuclear-spin chemistry can account for anomalously low observed
   NH$_2$-OPR values. We have tentatively explained OPR values larger
   than three by assuming that spin thermalization of NH$_2$ can
   proceed at least partially by H-atom exchange collisions with
   atomic hydrogen, thus increasing the OPR with decreasing
   temperature.  In this paper, we present quasi-classical trajectory
   calculations of the H-exchange reaction \mbox{NH$_2$ + H}, which
   show the reaction to proceed without a barrier, confirming that the
   H-exchange will be efficient in the temperature range of
   interest. With the inclusion of this process, our models suggest
   both that OPR values below three arise in regions with temperatures
   $\gtrsim20-25$~K, depending on time, and values above three but
   lower than the thermal limit arise at still lower temperatures.}
\maketitle
%
\section{Introduction}

Hydrides play a crucial role in astrochemistry as initial building
blocks of the chemistry in both diffuse and dense clouds. The new
spectroscopic window opened by the {\it Herschel} Space Observatory in
the submillimeter and in the far-infrared (FIR) has allowed the
detection of some fundamental and excited rotational transitions of
simple neutral or ionized hydrides in different types of sources and
especially towards the cold interstellar medium, in either the
envelopes of low- and high-mass star-forming regions or in more
distant regions along the lines-of-sight to these objects. In addition
to the detection of new hydrides such as ND \citep{bacmann2010},
\ce{HCl+} \citep{deLuca2012}, and \ce{ArH+} \citep{barlow2013}, doubly
and triply hydrogenated hydrides in their ortho and para forms such as
\ce{H2Cl+} \citep{lis2010}, \ce{H2O+}
\citep{ossenkopf2010,schilke2013,gerin2013}, \ce{NH2} and \ce{NH3}
\citep{hilyblant2010nh,persson2010,persson2012,persson2016}, and
\ce{H2O} \citep{emprechtinger2013} have been detected. Some of the
ortho-to-para ratios (OPRs) were found to be consistent with their
thermal values and some such as the cases of water
\citep{lis2013,flagey2013}, \ce{H3+} \citep{crabtree2011}, \ce{NH3}
\citep[][]{persson2012} and \ce{NH2} \citep{persson2016} were not.

Observing and being able to constrain OPRs in such environments can
bring crucial information about the prevailing physical conditions,
such as the temperature of the gas, and also, on the other hand, can
yield new constraints on the interstellar chemistry occurring in these
milieu.  For instance, the \ce{H2} OPR has been suggested as a
chemical clock in cold molecular gas
\citep{flower1984,flower2006op,pagani2009,pagani2011,pagani2013,brunken2014}.
Comprehensive analysis of how these OPRs arise involves a deep study
of the interstellar chemistry of these simple polyatomic species,
which is often poorly known, especially concerning the processes and
rates governing {\it (i)} the formation of ortho and para species and
{\it (ii)} the ortho-to-para conversion.

As an example, interest in the interstellar chemistry of
nitrogen-hydride species has arisen as a result of observations of the
lowest rotational transitions of the nitrogen hydrides \ce{NH},
\ce{NH2} and \ce{NH3} at far-IR wavelengths towards cold interstellar
sources
\citep{bacmann2010,hilyblant2010nh,persson2010,persson2012}. During
the last decades, different gas-phase and gas-grain models were
developed in an attempt to reproduce the observational data of these
hydrides but with at most moderate success
\citep{meyer1991,millar1991,wagenblast1993,oneill2002,hilyblant2010nh}.

Based on new theoretical and experimental data
\citep[\eg][]{flower2006n,hugo2009,honvault2011,honvault2012,rist2013,daranlot2013,daranlot2012,jorfi2009NO},
\cite{legal2014a} revised the understanding of nitrogen chemistry by
focussing on the study of the basic gas-phase processes for the
specific physical conditions of cold molecular gas.  They revisited
the low temperature kinetics of the nitrogen-bearing species and
provided a rigorous nuclear-spin chemistry of the N-hydrides
\citep{rist2013}. The result was a nearly pure gas-phase astrochemical
model, which does not consider any grain reactions except the
formation of \ce{H2} and charge exchange reactions. The network
includes nuclear-spin selection rules for the hydrogenated nitrogen
molecules and also for the hydrogen chemistry, which plays a crucial
role in the synthesis of nitrogen hydrides
\citep{lebourlot1991,dislaire2012}. This model was able to reproduce
the abundances and abundance ratios of \ce{NH}, \ce{NH2}, and \ce{NH3}
observed towards the envelope of the protostar IRAS 16293-2422. The
model was also the first to lead to an explanation of the puzzling
measurements of the OPRs of ammonia in cold diffuse gas ($T \sim
30$~K), which were found to be \citep[$\approx 0.5 - 0.7 \pm
0.1$,][]{persson2012,faure2013}, below the thermal value of unity.

\begin{figure}
 \centering
\resizebox{\hsize}{!}{\includegraphics{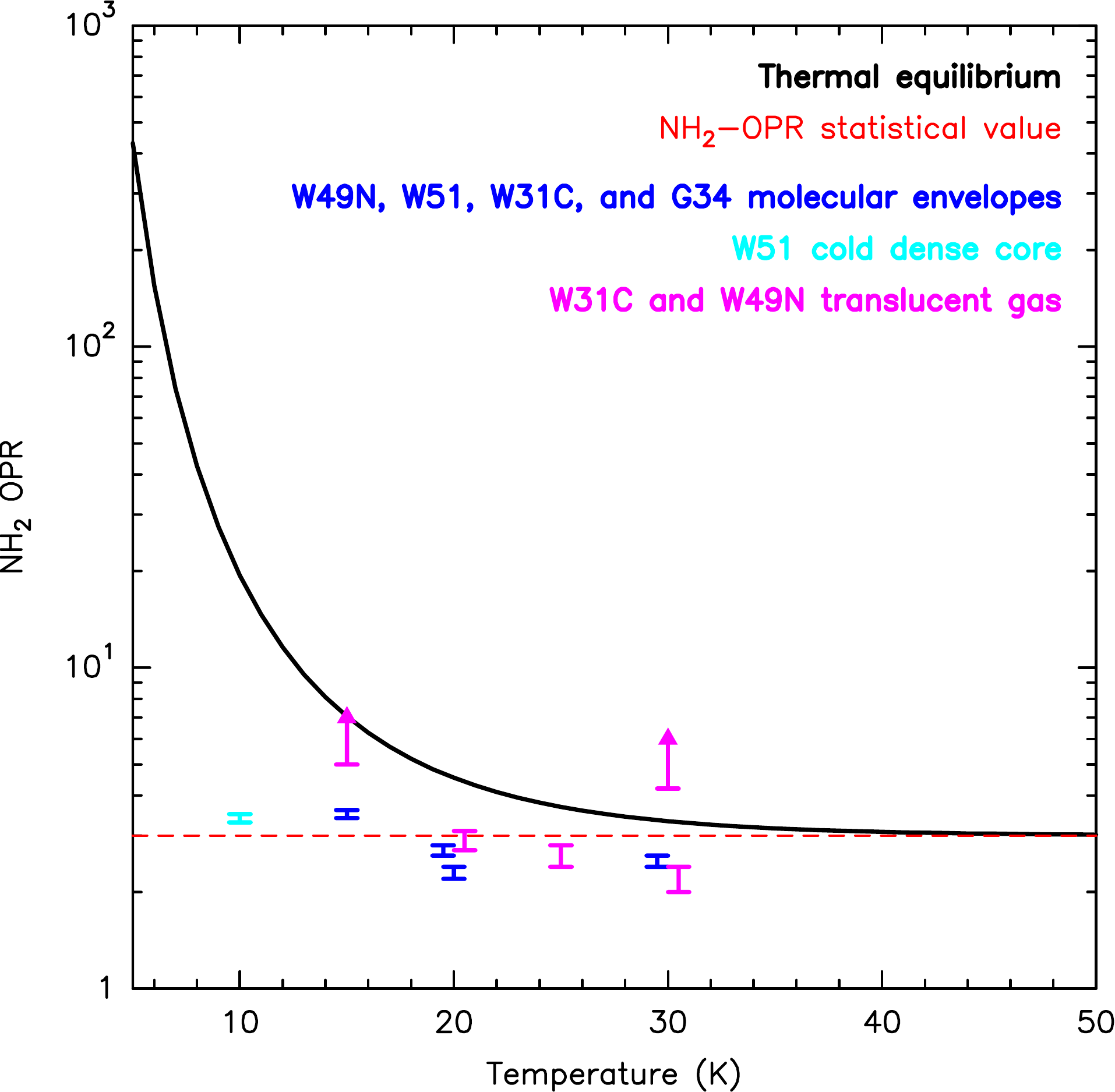}}
\caption{NH$_2$ OPR computed as a function of temperature at thermal
  equilibrium in black along with the observed OPR values from
  \cite{persson2016} in blue, cyan and pink at the lowest values of
  the observed temperature ranges \citep{persson2016}. For the sake of
  clarity the uncertainties of the observed temperatures are
  omitted. The NH$_2$-OPR statistical value of 3 is represented by the
  dashed red line.}
 \label{fig:NH2_OPR_ture_thermal_obs}
\end{figure}

\begin{figure}
 \centering
\resizebox{\hsize}{!}{\includegraphics{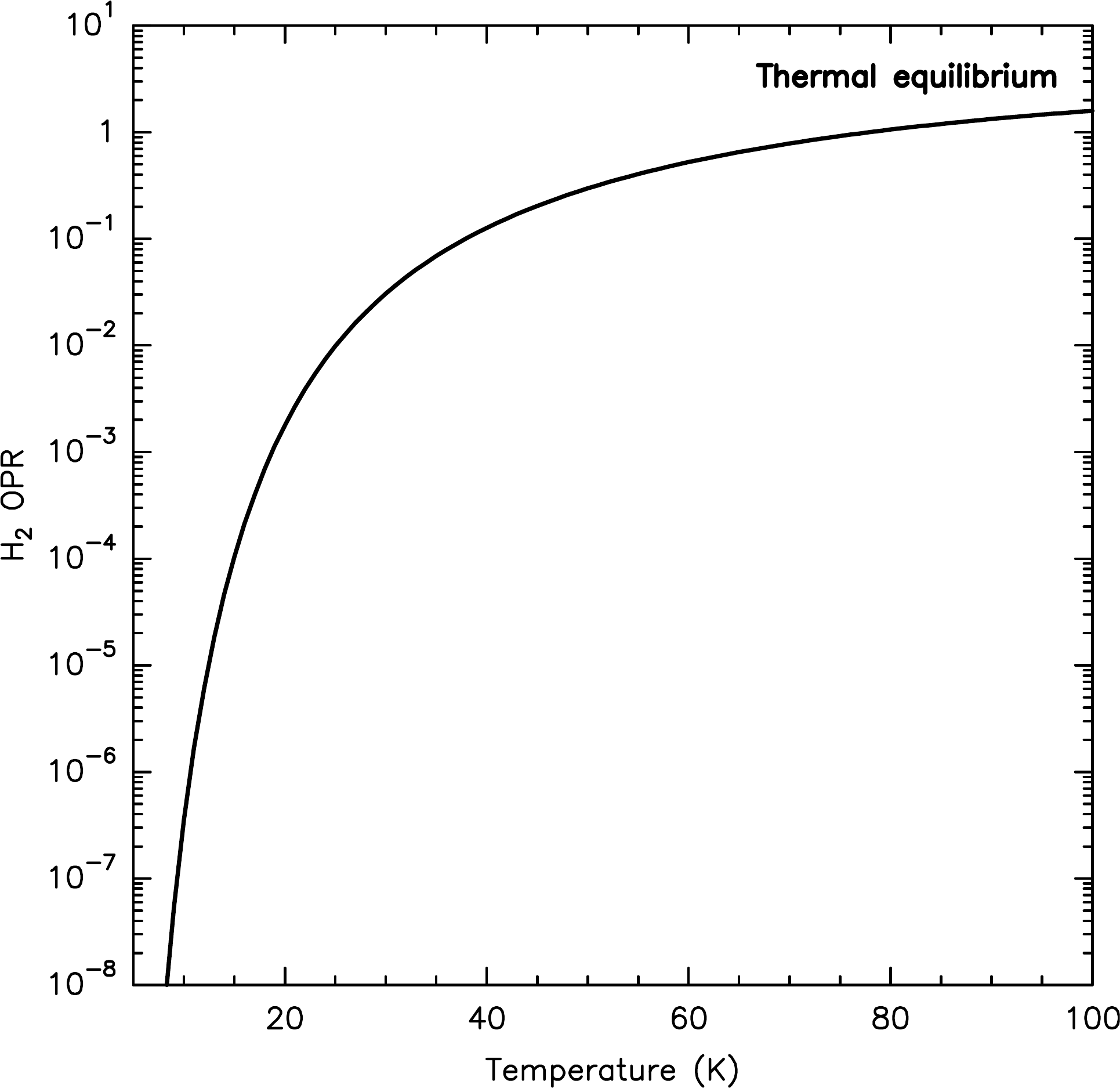}}
\caption{\ce{H2} OPR computed as a function of temperature at
  thermal equilibrium.}
 \label{fig:H2_OPR_ture_thermal}
\end{figure}

More recently, \cite{persson2016} investigated the non-thermal
observational values of the \ce{NH2} OPR measured towards the
high-mass star-forming regions W31C (G10.6-0.4), W49N (G43.2-0.1), W51
(G49.5-0.4), and G34.3+0.1 including translucent clouds in front of
these sources. These authors, focussing their study on the \ce{NH2}
OPR, did not investigate in detail the column densities of ortho and
para \ce{NH2}. The \ce{NH2}-OPR values observed by \cite{persson2016}
towards the different sources are displayed in
Figure~\ref{fig:NH2_OPR_ture_thermal_obs} with the lowest values of
the observed temperature ranges. The uncertainties in the observed
temperatures are very large; if the upper values were utilized in the
figure, the OPR ratio corresponding to these values should never
exceed the statistical value of three so that the OPR measurements
with values greater than three would likely have much larger
uncertainties than reported. Although the use of nuclear-spin
selection rules in an improved model leads to the reproduction of most
of the observed OPR values below three at reasonable temperatures, it
was necessary to find a mechanism that can at least partially
thermalize the OPR at particularly low temperatures where the thermal
OPR exceeds three and goes to infinity as the temperature goes to 0 K,
as shown in black in Figure~\ref{fig:NH2_OPR_ture_thermal_obs}. This
pattern occurs in the opposite sense from the H$_{2}$ thermal OPR,
which is depicted in Figure~\ref{fig:H2_OPR_ture_thermal}. The
difference arises because of the additional asymmetry of the ground
electronic state of NH$_{2}$.

This need led \cite{persson2016} to consider the poorly studied
NH$_{2}$ - H atom-exchange reaction as a mechanism to interconvert
NH$_{2}$ between its ortho and para forms, previously omitted in
models. With the assumption that this exchange occurs rapidly in both
directions, \cite{persson2016} were able to explain the large values
of the OPR observed in selected cold sources. But the H-exchange
reaction between H and NH$_2$ had not been studied in detail. Weak
experimental evidence from the saturated three-body reaction to
produce ammonia indicates at most a small barrier
\citep{pagsberg_pulse_1979}, while theoretical calculations indicate a
more substantial barrier and a conical intersection
\citep{mccarthy_dissociation_1987,zhu2012}.  In this paper, we present
a new theoretical calculation developed to determine if this reaction
is a plausible efficient para-to-ortho conversion pathway for
NH$_{2}$. We also determine how our calculated OPR is affected by
recent updates concerning gas-phase reactions between NH$_{2}$ and the
abundant oxygen and nitrogen atoms.

In Section~\ref{sec:OPR}, we discuss the OPR of \ce{NH2} at low
temperatures and how different non-thermal OPR values can be obtained,
while emphasizing the role of the \ce{H + NH2} H-exchange
reactions. Section 3 begins with a presentation of the theoretical
work developed in order to study the \ce{H + NH2} H-exchange reaction
proposed in \cite{persson2016}. This section continues with our
astrochemical results derived from six new nearly pure gas-phase
models, which differ from each other concerning: {\bf{\it (i)}} the
inclusion of new destruction reactions for NH$_{2}$, {\bf{\it (ii)}}
the initial form of hydrogen, {\bf {\it (iii)}} the cosmic-ray
ionization rate, {\bf {\it (iv)}} the gas density, and {\bf {\it (v)}}
the sulfur elemental abundance. In Section 4, we discuss the results,
chiefly our new findings and how they fit the observational data,
including the temperature range over which they can reproduce the
measured OPR of NH$_{2}$. Finally, Section 5 contains a summary of our
calculations and results.

\section{NH$_2$ OPR and ortho-para interconversion}
\label{sec:OPR}

\subsection{NH$_2$ OPR under thermal conditions}

The \ce{NH2} OPR under thermal conditions can be expressed by the equation:

\begin{equation}
{\rm OPR_{\ce{NH2}}}(T) = \frac{3 \sum^{ortho}_J \,g_J\,\exp(-E_{J_{K_a,K_c}}/k_BT)}{\sum^{para}_J \,g_J \,\exp(-E_{J_{K_a,K_c}}/k_BT)} 
\label{eq:fmuleROP}
\end{equation}
with $g_J$, the degeneracy of the total angular momentum, $E_J$ the
energy of the rotational levels, which also depends on the $K_a$ and
$K_c$ pseudo-quantum numbers corresponding to the projections of the
total angular momentum $J$ on the symmetry axes of the prolate and
oblate symmetric top limits, respectively
\citep{townes1955}. Ortho-\ce{NH2} corresponds to $K_a+K_c=2n$, and
para-\ce{NH2} to $K_a+K_c=2n+1$, with $n$ a non-negative integer.  For
simplicity the fine- and hyperfine-structure energies are omitted in
this formula.

Figure~\ref{fig:NH2_OPR_ture_thermal_obs} shows in black the variation
of the OPR of NH$_2$ as a function of the temperature at thermal
equilibrium. At high temperatures, where many rotational levels are
populated, the thermal OPR is equal to three, the ratio of the
statistical weights of all ortho and para levels. At very low
temperatures or strongly subthermal rotational excitation, only the
lowest ortho and para rotational states are populated. Due to the
anti-symmetry of the ground electronic wave function, the ground
rotational-spin state ($0_{00}$) is an ortho state of NH$_2$ while the
lowest para rotational state ($1_{01}$) lies 30.4 K higher
\citep{persson2016}. Thus, with the additional assumption that the
fine-structure and hyperfine-structure energies are degenerate, at low
temperatures, the \ce{NH2} OPR can be expressed by the equation:

\begin{centering}
  \begin{eqnarray} 
    {\rm OPR_{\ce{NH2}}} (T_{\rm low}) \thickapprox
    \frac{3 \, g_{J=0}\exp(-E_{0_{00}}/T)}{g_{J=1}\exp(-E_{1_{01}}/T)
    }=\exp\left(\frac{-\Delta E}{T}\right) \nonumber \\
    =\exp\left(\frac{30.4}{T}\right)
 \label{eq:NH2-OPR-approx_lowT},
\end{eqnarray}
\end{centering}
where $\Delta E = E_{0_{00}} - E_{1_{01}} = -30.4~\rm{K}$ is the
energy difference between the two ground rotational-spin states. Thus,
the low-temperature OPR continues to increase strongly with decreasing
temperature.

In the interstellar medium, true thermodyamic equilibrium, at least
between kinetic and rotational energy, is only reached at the higher
densities attainable.  At lower densities, the rotational excitation
can be subthermal, so that the rotational temperature lies below the
kinetic temperature.  In this case, it would be more appropriate to
use the rotational temperature in the OPR formulae.  For regions in
which NH$_{2}$ is detected in absorption, as is the case for the
observed values considered in the present study, the low temperature
limit is normally adequate.

\subsection{NH$_2$-OPR values below thermal equilibrium}
\label{subsec:below_thermal_value}

In \cite{persson2016}, it was suggested that those observed
\ce{NH2}-OPR values lower than the thermal value could arise because
in such low temperature environments \ce{H2} is para-enriched. The
H$_2$ OPR controls the key initiating reaction involved in the
formation of nitrogen hydrides and in particular the formation of the
ammonium ion, NH$^+_4$, the main direct precursor of NH$_2$ in cold
dense gas \citep{persson2016}. NH$_2$ can also be produced through the
dissociative recombination of \ce{NH3+} with electrons. This pathway
is not dominant for cold dense gas but can become more efficient for
diffuse and translucent gas, where the electron fraction is
higher. For dense cold gas conditions, the nuclear spin branching
ratios in the dissociative recombination with electrons of the three
spin configurations of NH$^+_4$ (ortho, meta, and para) primarily
determines the NH$_2$ OPR if it is only due to formation processes,
according to the formula:

\begin{centering}
\begin{equation}
\rm OPR_{\ce{NH2} \, formation}=\frac{2 \times MPR_{NH^+_4} \, + \,
 \frac{4}{3} \times OPR_{NH^+_4} \, + \, 1}{\frac{2}{3} \times OPR_{NH^+_4} \, + \,
 1}
\label{eq:opNH2-formation}
\end{equation}
\end{centering}
where MPR stands for meta-to-para ratio. Considering these
nuclear-spin selection rules, the gas-phase spin-conservation model
developed in \cite{legal2014a,legal2014b} was able to reproduce the
\ce{NH2}-OPR values below the statistical value of 3:1 observed
towards the molecular envelopes of W31C, W51 and G34.3, and in
translucent gas towards W31C. However, as mentioned in the
Introduction, this model was not able to reproduce a variety of
\ce{NH2}-OPR values above three found towards the molecular envelope
of W49N, a dense filament connected to W51, and some translucent gas
towards W31C and W49N.

\subsection{Plausible H-exchange reaction between o-NH$_2$ and p-NH$_2$}
\label{subsect:H-exchange_reaction}

In order to understand the \ce{NH2}-OPR values found above the
statistical value of three, \citet{persson2016} suggested that once
formed, the NH$_2$ ortho and para radicals should undergo an
H-exchange reaction with H, allowing interconversion between the
lowest rotational states of ortho-NH$_2$ and para-NH$_2$, hereafter
\ce{o-NH2} and \ce{p-NH2} respectively:
\begin{equation}
\begin{centering}
\ce{p-NH_2 + H <=> o-NH_2 + H + $30.4$ K},
\end{centering}
\end{equation}
Such processes are likely to thermalize the OPR given sufficient
time. But if the reactive collisions are inefficient, either because
they are inherently slow or because there are faster competitive
destruction mechanisms, the OPR should lie in between the formation
value of the NH$_2$ OPR, produced by exothermic dissociative
electronic recombination of NH$^+_4$, and the thermalized value, \eg\
7.6 at 15~K or 21 at 10~K. If, on the other hand, the average time
between two successive ortho/para exchange collisions between H and
\ce{NH2} is negligible compared with the average lifetime of \ce{NH2}
then the NH$_2$ OPR should reflect the temperature of the gas and
follow the LTE OPR, which can be quite high at sufficiently low
temperatures.

To quantify these points, \cite{persson2016} added the following two
reactions to the \cite{legal2014a} model:

\begin{equation}
\ce{ H + o-NH_2 ->[k_{\rm{o \rightarrow p}}]  H +
  p-NH_2$,$}\label{eq:conversion_onh2_pnh2}
\end{equation}

\begin{equation}
\ce{ H + p-NH_2 ->[k_{\rm{p \rightarrow o}}]  H + o-NH_2$,$}\label{eq:conversion_pnh2_onh2}
\end{equation}

\noindent
with $k_{\rm{o \rightarrow p}}=k_{\rm{p \rightarrow o}}\exp(-30.4/T)
\cccs$, which should be accurate at low temperatures. For the
$k_{\rm{p \rightarrow o}}$ rate coefficient, a typical radical-radical
value of $1\tdix{-10}\cccs$ was chosen initially.

\begin{figure}
 \centering
\resizebox{\hsize}{!}{\includegraphics{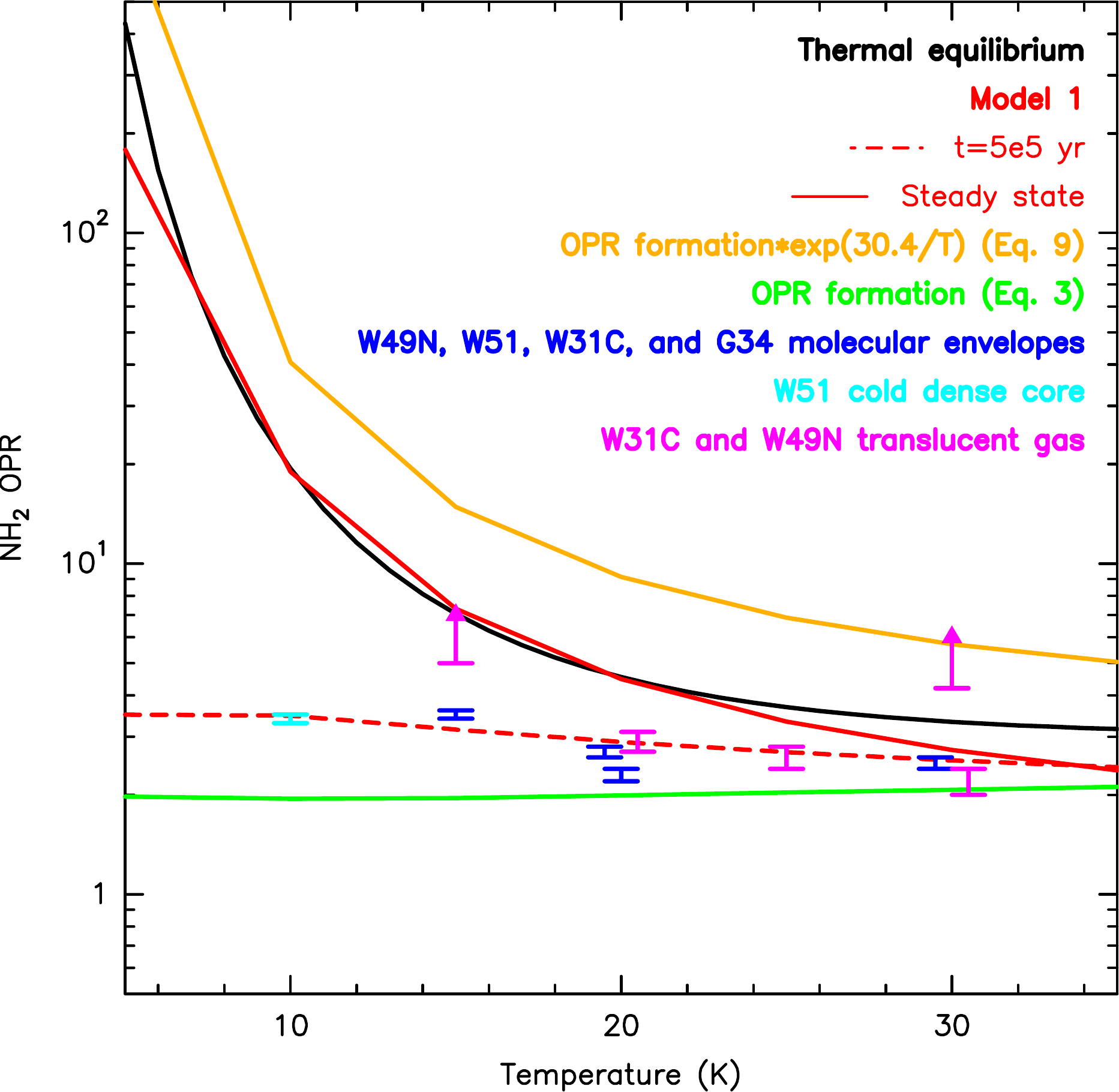}}
\caption{\ce{NH2} OPR as a function of temperature computed: {\it (i)}
  at thermal equilibrium, {\it (ii)} with Model 1 at steady state and
  an earlier time which best fits the observed values, and {\it (iii)}
  with the analytical
  formulae~(\ref{eq:opNH2-formation})~and~(\ref{eq:opNH2-therm}). The
  details of Model 1 are described in
  Sect.~\ref{subsect:H-exchange_reaction}. The observed OPR values
  from \cite{persson2016} are represented within their error bars in
  blue, cyan and pink at the lowest values of the observed temperature
  ranges \citep{persson2016}. For the sake of clarity the
  uncertainties of the observed temperatures are omitted.}
 \label{fig:NH2_OPR_ture_thermal_model1_analytical}
\end{figure}

We label, as Model 1, the model b used in \cite{persson2016}, which
takes into account the network of reactions of \cite{legal2014a} with
the addition of the forward and backward \ce{NH2 + H}
reactions~(\ref{eq:conversion_onh2_pnh2})
and~(\ref{eq:conversion_pnh2_onh2}). The physical fixed conditions
used to run this model are typical of dense gas: a density $n_{\rm H}
= 2\tdix{4}\ccc$, the commonly used value $\zeta=1.3 \times
10^{-17}\pers$ for the cosmic-ray ionization rate
\citep{spitzer1968,prasad1980a,wakelam2005,vastel2006} and a visual
extinction of 10~mag. The initial abundances are those used in
\cite{persson2016}. In particular for the C/O and sulfur elemental
abundance, \cite{persson2016} chose values that best fit the observed
N-hydride abundances: 0.6 and $3\tdix{-6}$, respectively. Despite the
fact that dense gas models usually consider a higher depleted sulfur
elemental abundance, this value remains poorly constrained in dense
gas. However, lately the use of such a high value has been required to
fit observations in the Horsehead nebula \citep{goicoechea2006} and
even more recently in the Barnard 1 region \citep{fuente2016}. The
modifications leading to other models will be discussed later. With
observed temperatures assumed to be near the bottom of the observed
temperature ranges \citep{persson2016}, Model 1 reproduces the
NH$_2$-OPR values observed above three at appropriate times and
temperatures \citep{persson2016}, as shown in
Figure~\ref{fig:NH2_OPR_ture_thermal_model1_analytical} by the dashed
red line. The range of times is less than the time to steady state.

The model results at steady state can be analytically understood as
follows. The evolution of the \ce{o-NH2} and \ce{p-NH2} abundances can
be expressed by the following simple kinetic equations:

\begin{gather}
\frac{d[\ce{o-NH2}]}{dt} = F_o -[\ce{o-NH2}]D_{o}\\
\frac{d[\ce{p-NH2}]}{dt} = F_p - [\ce{p-NH2}]D_{p},
\label{eq:NH2-ab_evol}
\end{gather}
where $F_{o}$ and $F_{p}$ are the formation rates and $D_{o}$ and
$D_{p}$ the effective destruction rate coefficients for o-NH$_{2}$
and p-NH$_{2}$, respectively. With the assumption that the
destruction rate coefficients are dominated by the H-exchange reactions,
we obtain at steady state that

\begin{centering}
\begin{equation}
{\rm OPR_{\ce{NH2}}} ={\rm OPR_{\ce{NH2} \, formation}} \times \exp(30.4/T).
\label{eq:opNH2-therm}
\end{equation}
\end{centering}

The \ce{NH2} OPR obtained from the formation rates, expressed in
eq.~(\ref{eq:opNH2-formation}), does not have any temperature
dependence in the temperature range considered \citep{persson2016}, as
displayed in green in
Figure~\ref{fig:NH2_OPR_ture_thermal_model1_analytical}. Thus, the
simple analytical expression of equation~(\ref{eq:opNH2-therm}) for
the \ce{NH2} OPR at steady state reflects the increase with decreasing
temperature as also seen in Model 1. Indeed, the analytical expression
for the \ce{NH2} OPR runs parallel to but somewhat above the thermal
equilibrium result as represented in mustard yellow in
Figure~\ref{fig:NH2_OPR_ture_thermal_model1_analytical}, while the
Model 1 result at steady state, in the red solid line, is almost
identical to the thermal equilibrium result in black over a wide
temperature range. A more detailed discussion of an improved
analytical approximation with other destruction processes for the
ortho and para forms of NH$_{2}$ can be found later in the text. Other
destruction processes will lessen the importance of the NH$_{2}$ + H
ortho-para exchange reactions and reduce the calculated OPR values.

The significant impact of the \ce{H + NH2} H-exchange reactions on the
\ce{NH2} OPR at low temperatures has led us to investigate the
reactions from a theoretical point of view. In the next section, we
discuss our calculations of the rate coefficients.

\section{Results}

\subsection{Theoretical calculations of the \ce{H + NH2} H-exchange
  reaction-rate coefficients}
\label{theoretical_calc}

\begin{figure}
 \centering
\resizebox{\hsize}{!}{\includegraphics[angle=-90]{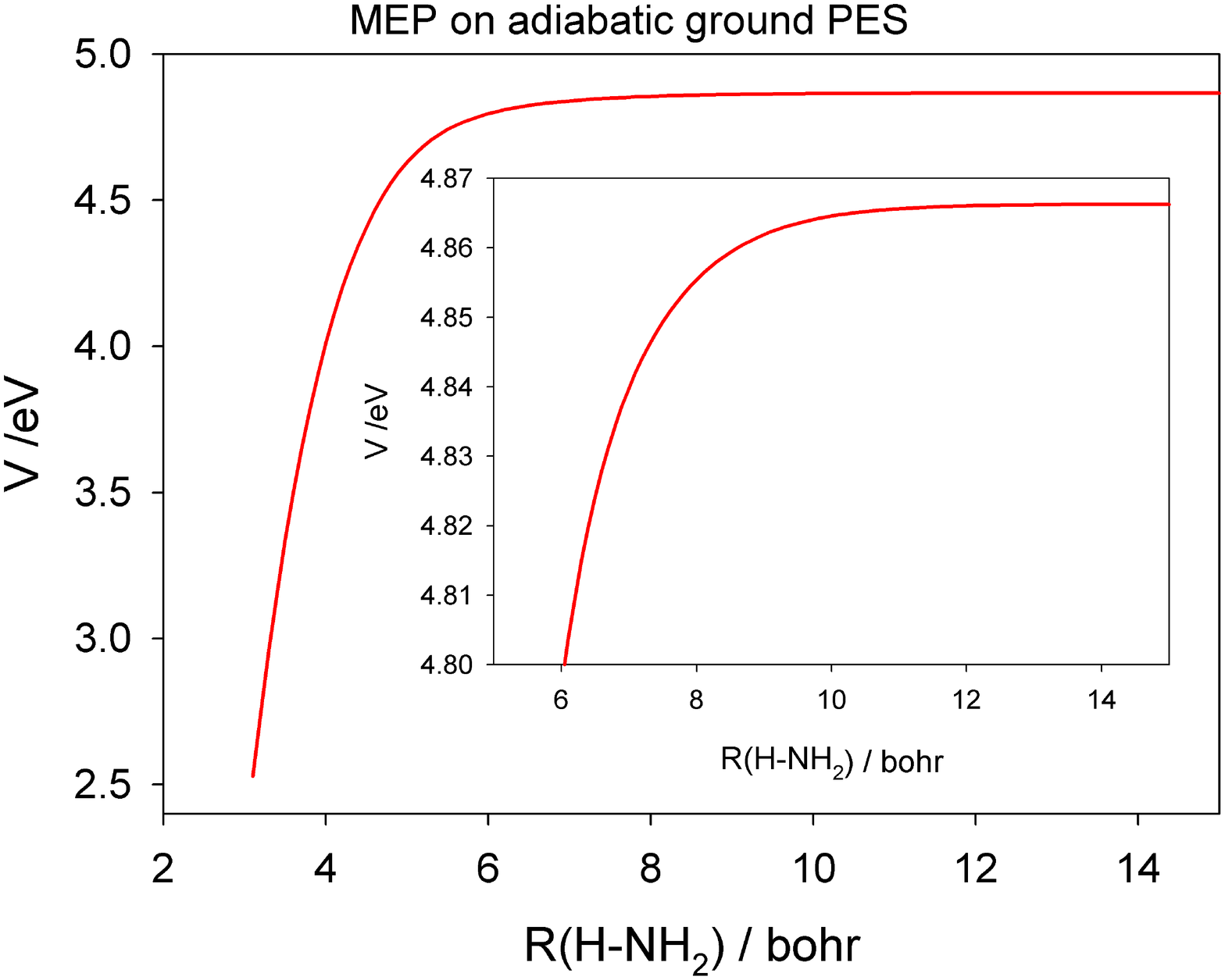}}
\caption{Minimum energy path (MEP) from the \ce{H + NH2} asymptote on the
  ground potential energy surface (PES) of \ce{NH3}.}
 \label{fig:Vmep-H+NH2}
\end{figure}

It is impossible to simulate directly the conversion between the ortho
and para forms of \ce{NH2} using classical mechanics because it does
not recognize symmetry of the wave function, yet a quantum mechanical
scattering calculation is still too expensive numerically. In this
work, we provide an estimate of the low-temperature rate coefficient
for the exchange reaction \ce{H^' + NH2 -> NHH^' + H}, serving as a
proxy of the ortho-para conversion. The calculations were carried out
using a modified quasi-classical trajectory (QCT) method
\citep{hase1998} as described below. The potential energy surface
(PES) used in the QCT calculation is from that devised by Zhu and
Yarkony (ZY) and their coworkers \citep{zhu2012} which is a 2x2
quasi-diabatic permutation invariant potential matrix designed for
studying the photodissociation of ammonia (\ce{NH3}) in its first
absorption band. Quantum dynamical studies on these PES have achieved
excellent agreement with experimental data, validating the PES
\citep{ma_2012,xie2014}. Only the lower adiabat, which corresponds to
the ground electronic state of \ce{NH3}, was used in our calculations
reported here. In Figure \ref{fig:Vmep-H+NH2}, the minimum energy path
from the \ce{H + NH2} asymptote is displayed and clearly there is no
barrier.

A major deficiency of the QCT method is the possible violation of
zero-point energy (ZPE). This problem is especially severe at low
temperatures, because ZPE represents a large fraction of the total
energy. In this work, the ZPE effect is approximately dealt with using
a scheme newly proposed by Hase and coworkers \citep{paul2016}. When a
trajectory exits the strongly interacting region, the vibrational
energy of the \ce{NH2} is calculated. If the energy is less than the
ZPE, the momenta of all atoms in the system are reversed and the
trajectory is forced back to the strongly interacting region without
violating energy conservation. Only those trajectories with \ce{NH2}
internal energies larger than the ZPE are allowed to dissociate.

QCT calculations in this work were implemented in VENUS
\citep{hu1991}. The trajectories were initiated with a 9.0~\ang\
separation between reactants, and terminated when products reached a
separation of 6.0~\ang. The ro-vibrational energies of the \ce{NH2}
reactant and relative translational energies were sampled from the
Boltzmann distribution at a specific temperature. The propagation time
step was selected to be 0.05~fs. Trajectories were discarded if (a)
the propagation time reached 100~ps in each interval of two
consecutive momentum reversing operations, (b) the number of momentum
reversing exceeds 100, or (c) the total energy failed to converge to
0.05~kcal/mol. The maximal impact parameters ($b_{max}$) were 8.2 and
7.8~\ang\ for 50 and 100~K, respectively. A total of 1144 (2117)
trajectories were calculated at 50 (100)~K. The thermal rate
coefficient is computed by the following expression:

\begin{equation}
k(T)=\left(\frac{8k_BT}{\pi\mu}\right)^{1/2} \pi b^2_{max} \frac{N_r}{N_{tot}}
\end{equation}
where $N_r$ and $N_{tot}$ are the numbers of reactive and total
trajectories and $\mu$ is the reduced mass of reactants.

Table ~\ref{tab:calculation_results} lists the calculated rate
coefficients for the \ce{H + NH2} H-exchange reaction with/without the
ZPE constraint. For the results with the ZPE constraint, the
statistical errors at two temperatures are somewhat large due to
smaller number of accepted trajectories. It can be noticed that the
results with the ZPE constraint are much smaller than those without
the ZPE constraint.  The exchange reaction proceeds via a
complex-forming mechanism, and trajectories spend most of the time in
the \ce{NH3} potential well. The small rate coefficients reflect, at
least partially, the very inefficient energy randomization in the
\ce{NH3} complex.

\begin{table}
\centering
\caption{Rate coefficients of the H+NH$_{2}$ exchange reaction
  with/without the ZPE constraint.}
\begin{tabular}{ccc}
  \hline\hline
  Temperature &\multicolumn{2}{c}{Rate Coefficient $k$}\\
  (Kelvin) &\multicolumn{2}{c}{($\dix{-10}$ \cccs)}\\
  \hline 
  &Without ZPE constraint&With ZPE constraint\\
  \hline 
  50	&3.583 (4.0\%)&	0.800 (15.3\%)\\
  100	&4.179 (1.9\%)&	2.253 (7.4\%)\\
  \hline 
\label{tab:calculation_results}
\end{tabular}
\end{table}

Concerning the rate coefficients computed and presented in
Table~\ref{tab:calculation_results} one can wonder if these
theoretical calculations are still valuable at temperatures below 50~K
since interstellar temperatures in the regions of interest for the
\ce{NH2} observations range down to 10~K with the lower temperatures
the more interesting. Especially one can wonder how rapid the \ce{NH2
  + H} reaction might be at such low temperatures, given that the ZPE
problem might be worse. Due to the computational costs, the
calculations at lower temperatures (\ie $< 50$~K) would have been even
more demanding, as fewer and fewer reactive trajectories can be
found. So we did not think it worthwhile to do it because all that we
need is an estimate of the order of magnitude, and the data at 50~K
should be able to provide that. Thus, as a result, these theoretical
calculations clearly show that {\it (i)} there is no barrier for the
\ce{H + NH2} H-exchange reaction as one can see in
Figure~\ref{fig:Vmep-H+NH2} and {\it (ii)} our calculated coefficient
($\sim \dix{-10} \cccs$) is consistent with that chosen in
\cite{persson2016}.

\subsection{Impact of the H-exchange rate coefficient on the \ce{NH2}
  OPR}

\begin{figure}[!th]
 \centering
\resizebox{\hsize}{!}{ \includegraphics{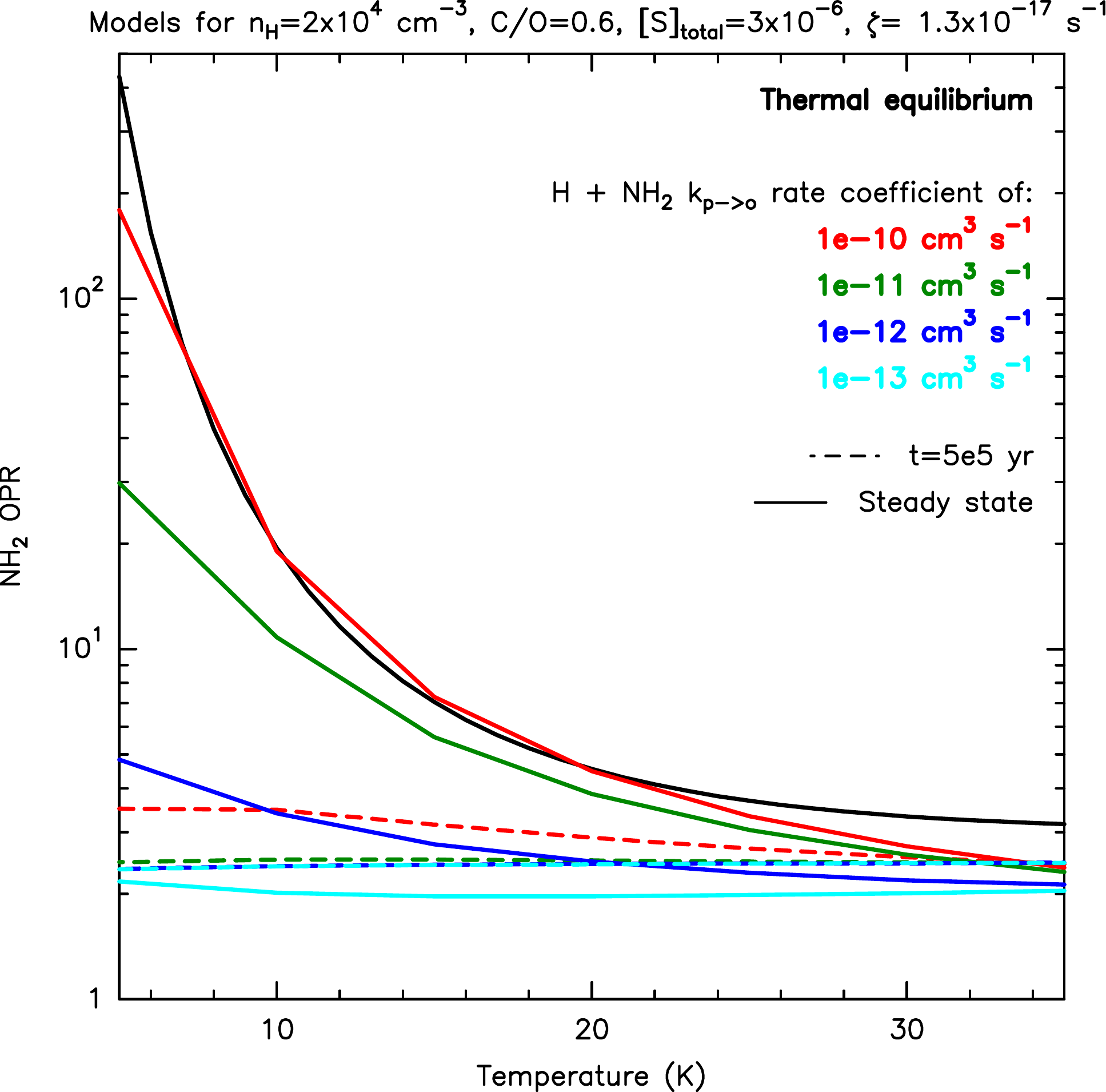}}
\caption{Impact of the $k_{\rm p \rightarrow o}$ rate coefficient of
  the \ce{H + NH2} H-exchange reaction on the NH$_2$ OPR as a function
  of temperature computed with Model 1 for the physical conditions
  given in the super title at steady state and an earlier time. The
  details of Model 1 are described in
  Sect.~\ref{subsect:H-exchange_reaction}.
}
\label{fig:NH2-ROP_ture_4rates}
\end{figure}

The theoretical calculations of the \ce{H + NH2} H-exchange rate
coefficient discussed in Sect.~\ref{theoretical_calc} have shown that
the use of $1\tdix{-10}\cccs$ for $k_{\rm p \rightarrow o}$ in Model 1
is realistic, although there are uncertainties in the calculation
involving the role of the ZPE. In order to study the impact of
lowering this rate coefficient on the efficiency of the the \ce{H +
  NH2} H-exchange reaction, we present in this section the results
when the $k_{\rm{p \rightarrow o}}$ rate coefficient is gradually
lowered. In Figure~\ref{fig:NH2-ROP_ture_4rates}, the results are
plotted for the values $1\tdix{-10}\cccs$ (red lines)
$1\tdix{-11}\cccs$ (dark green lines), $1\tdix{-12}\cccs$ (blue
lines), and $1\tdix{-13}\cccs$ (cyan lines). We see that decreasing
the \ce{H + NH2} H-exchange reaction rate coefficient until
$1\tdix{-13}\cccs$ in Model 1 gradually lowers the increase in the OPR
as temperature is decreased.

By a certain rate coefficient, here $1\tdix{-13}\cccs$, the model can
no longer produce OPR values above 3 at the lowest temperatures
considered. Also for most values of the exchange rate coefficient, the
OPR values calculated at a time earlier than steady state are lower
than at steady state although this is not the case for the lowest rate
coefficient. Thus, even if $k_{\rm p \rightarrow o}$ is several orders
of magnitude lower than the calculated value, OPR values greater than
three can still be obtained, at least at steady state, at
astronomically meaningful temperatures, and used to explain the high
OPR values from \cite{persson2016} depicted in
Figures~\ref{fig:NH2_OPR_ture_thermal_obs} and
~\ref{fig:NH2_OPR_ture_thermal_model1_analytical}.

\subsection{Impact of chemical updates on the \ce{NH2} OPR and
    abundances}

\label{subsect:updates}

\begin{figure}
 \centering
 \resizebox{\hsize}{!}{ \includegraphics{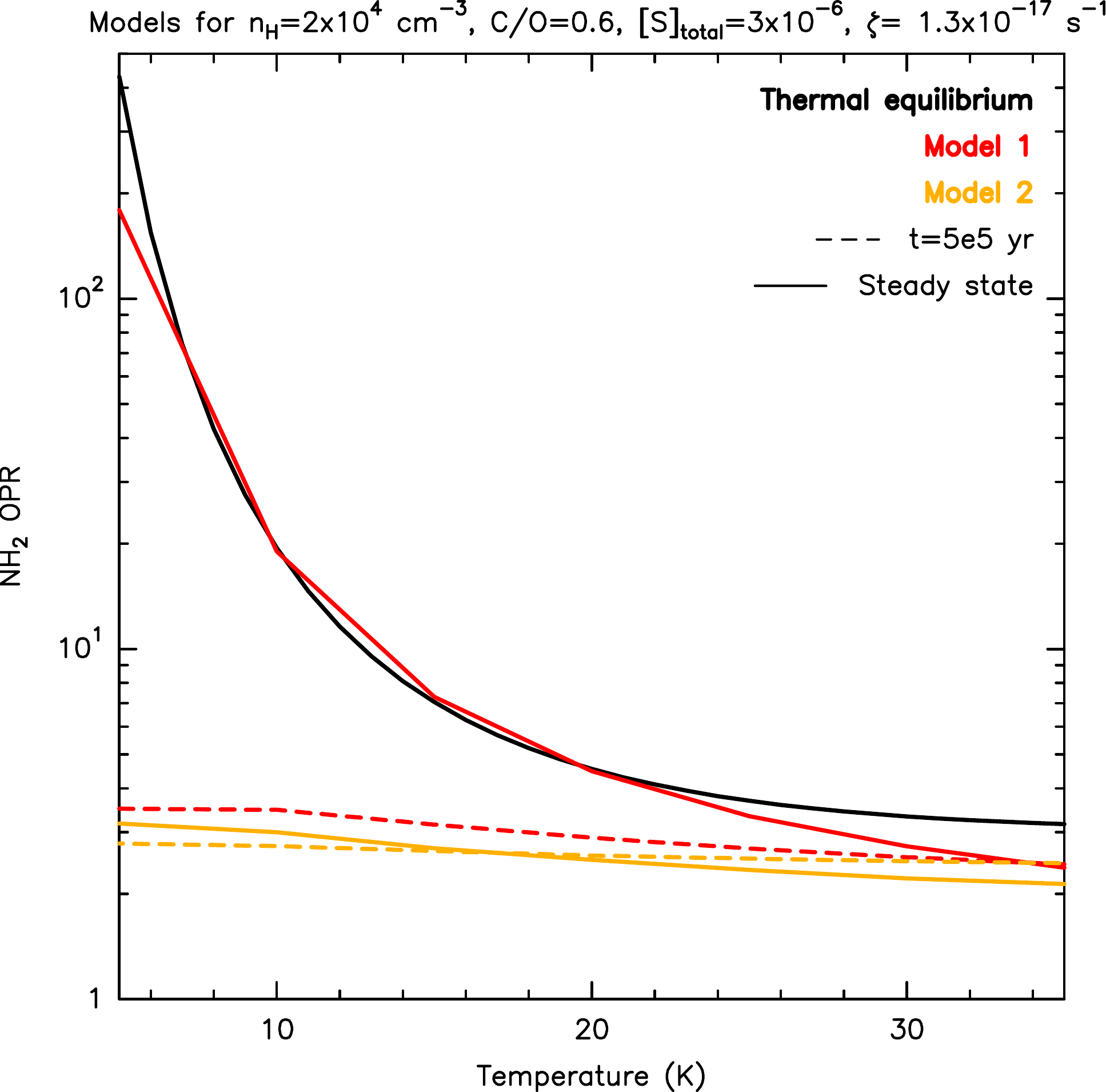}}
 \caption{OPR of NH$_2$ computed as a function of temperature for
     the physical conditions given in the super title at thermal
   equilibrium, and at steady state and an early-time for Models 1 and
   2. The details of the models can be found in
   Sect.~\ref{subsect:H-exchange_reaction} and ~\ref{subsect:updates},
   respectively. Note again that the Model 1 result at steady state
   lies close to the equilibrium result over a wide range of
   temperatures.}
\label{fig:NH2-ROP_ture}
\end{figure}

In this section we present the results obtained with a new model,
Model 2, which contains new or revised destruction processes for
NH$_{2}$ and NH. For NH$_{2}$, we include a new destruction pathway
with atomic nitrogen and a revision of the destruction rate
coefficient with atomic oxygen, while for NH, we update several rate
coefficients for its destruction \citep{wakelam2013}. This model is
represented in mustard yellow in Figure~\ref{fig:NH2-ROP_ture}, which
contains a plot of the OPR for NH$_{2}$ as a function of temperature
for an early time and at steady state. The new and revised rate
coefficients for \ce{NH2} and \ce{NH} destruction are presented in
Table~\ref{tab:NH2_NH}.

\begin{table*}[\!htb]
\centering
\tiny
\caption{\ce{NH2} and NH destruction updates.}
\begin{tabular} {llllllcccl} 
 \hline\hline
\multicolumn{6}{l}{Chemical
  reactions$^{(a)}$}&$\alpha$&$\beta$&$\gamma$&References\\
&&&&&&(\cccs)\\
\hline
\ce{NH2}&     N&       $\rightarrow$ &\ce{N2}&      H&       H&                  1.2(-10)&   0.00&   0.00&KIDA$^{(b)}$\\
\ce{NH2}&     O&       $\rightarrow$&NH&      OH&       &                 7.0(-12)&  -0.1&   0.00& KIDA$^{(c)}$\\
&      &        &        &      &                        &              3.5(-12)&   0.5&   0.00&\cite{legal2014a}$^{(d)}$ \\
\ce{NH2}&     O&       $\rightarrow$&HNO&     H&     &                     6.3(-11)&  -0.1&   0.00& KIDA$^{(c)}$\\
\ce{NH2}&     O&       $\rightarrow$&NO&    \ce{H2}&     &                     0.00&   0.00&   0.00& KIDA$^{(c)}$\\
NH&      N&       $\rightarrow$ &\ce{N2}&      H&    &                     5.0(-11)&   0.1&   0.00&KIDA$^{(e)}$\\  
NH&      O&       $\rightarrow$ &OH&      N&    &                     0.00&   0.00&   0.00&KIDA$^{(e)}$\\  
&      &        &        &      &                        &  2.9(-11)&   0.00&   0.00&\cite{legal2014a}$^{(d)}$ \\  
NH&      O&       $\rightarrow$ &NO&      H&&                         6.6(-11)&   0.00&   0.00&KIDA$^{(e)}$\\   
\hline 
\label{tab:NH2_NH}
\end{tabular}
\begin{list}{}{}
  \item\textbf{Notes}: Numbers in parentheses are powers of 10.
  \item $^{(a)}$ For the reactions involving \ce{NH2} as a reactant,
    the same rate coefficient is used for both ortho and para forms.
  \item $^{(b)}$ \cite{wakelam2013}, KIDA datasheet (http://kida.obs.u-bordeaux1.fr/datasheet/datasheet\_5734\_N+NH2\_V1.pdf);
  \item $^{(c)}$ \cite{wakelam2013}, KIDA datasheet
    (http://kida.obs.u-bordeaux1.fr/datasheet/datasheet\_290\_O+NH2\_V4.pdf);
  \item $^{(d)}$ from \cite{prasad1980a};
  \item $^{(e)}$ \cite{wakelam2013}, KIDA datasheet (http://kida.obs.u-bordeaux1.fr/datasheet/datasheet\_1500\_O+NH\_V7.pdf).
  \end{list}
\end{table*}

The \ce{N + NH2} reaction, which was not included in previous models
\citep{legal2014a,persson2016}, was experimentally studied by
\cite{whyte1983} and \cite{dransfeld1987} and has been shown to have
one main efficient exothermic production channel: \ce{N2 + H + H}
\citep{whyte1984}.  So this reaction adds an \ce{NH2} destruction
pathway with a temperature-independent rate coefficient of
$1.2\tdix{-10}\cccs$. The updated rate coefficient for the \ce{NH2 +
  O} reaction increases the destruction rate by a factor of~20. With
these enhancements, the new and updated destruction rate coefficients
for NH$_{2}$ affect the temperature dependence of the OPR by competing
with the {H + NH$_{2}$} H-exchange reaction and reducing its effect,
as represented in Figure~\ref{fig:NH2-ROP_ture} where we compare Model
2 with Model 1, which contains less effective destruction rates for
NH$_{2}$, at two different timescales. In this figure, it can be noted
that the updates highly affect the steady-state results, for which the
thermalization processes is strongly inhibited. Consequently, Model 2
can barely lead to OPR values greater than three at low temperatures.

From an analytical point of view, equation~(\ref{eq:opNH2-therm}) must
now include these destruction processes, and thus becomes:

\begin{centering}
  \begin{equation} {\rm OPR_{\ce{NH2}}} = {\rm OPR_{\ce{NH2} \,
        formation}} \times \frac{k_{\rm{p \rightarrow o}}[\rm
      H]+k_{\rm O}[O]+k_{\rm N}[N]}{k_{\rm{o \rightarrow p}}[\rm
      H]+k_{\rm O}[O]+k_{\rm N}[N]}
\label{eq:opNH2-dependence2},
\end{equation}
\end{centering}

\noindent
where the new rate coefficients refer to the sum of the reaction
channels. With the updated NH$_{2}$ destruction rates by N and O,
these processes are no longer negligible compared with the \ce{H +
  NH2} H-exchange reaction since these updated rates are the same
order of magnitude as the H-exchange reaction for two reasons. First,
the rate coefficients are $\approx 1\tdix{-10}\cccs$ and, in addition,
the abundances of N and O remain larger than that of H, by factors of
$\sim 2$ and $\sim 5$ respectively, until $\sim \dix{6}$~yr, as shown
in Figure~\ref{fig:H_N_O_NH2_ab}, by the comparison of Model 1 (our
old model) and Model 2, which contains the \ce{NH2} destruction
updates. So we need to find a new way to efficiently thermalize the
\ce{NH2} OPR at the lower temperatures.

From Figure~\ref{fig:H_N_O_NH2_ab}, it can also be noticed that at
20~K the abundances of both ortho and para \ce{NH2} are decreased by
the updates of the \ce{NH2} destruction reactions. At steady state,
these decreases are factors of $\sim130$ and $\sim70$, respectively,
while at more physical early times such as $\dix{6}$~yr, the decreases
are $\sim50$ and $\sim30$, and at $\dix{5}$~yr, they are $\sim2.4$ for
both ortho and para cases. Thus, the predicted overall abundance of
\ce{NH2} is decreased, but the effect is only large at long
times. Rather than compare the reduced total \ce{NH2} abundance with
observed values here in more detail, we prefer to delay this
comparison to a future paper in which we will utilize a gas-grain
chemical network more appropriate for total abundances.

\begin{table*}
\centering
\caption{Different models used in this work\tablefoottext{a}.}
\renewcommand{\arraystretch}{1.4}
\begin{tabular}{|l|c|c|c|c|c|c|c|c|}
 \hline
\multicolumn{1}{|c|}{\backslashbox[70mm]{Modifications}{Models}}& 1& 2&3& 4&5&6&7&1'\\
\hline 
\ce{H + NH2} H-exchange addition (reactions~\ref{eq:conversion_onh2_pnh2} and ~\ref{eq:conversion_pnh2_onh2})& X & X & X&X &X&X &X &X\\
\hline
\ce{NH2} destruction updates (see Table~\ref{tab:NH2_NH})& & X & X&X &X&X &X&\\
\hline
$\rm[H_{tot}]_{ini}=2 \times [\ce{H2}]$& X &X & &&X&X &&X\\
\hline
$[\rm H_{tot}]_{ini}=[H] $&  & & X&&&&&\\
\hline
$[\rm H_{tot}]_{ini}= \frac{1}{2}\times [H] + [\ce{H2}]$&  & &
&X&&&X&\\
\hline
$\zeta=1.3\tdix{-17}\pers$& X&X &X &X&&&&\\ 
\hline
$\zeta=3\tdix{-17}\pers$&  & & &&X&&&\\ 
\hline
$\zeta=2\tdix{-16}\pers$&  & & &&&X&X&X\\ 
\hline
$n_{\rm H}=2\tdix{4}\ccc$&X  &X & X&X&X&&&\\ 
\hline
$ n_{\rm H}=1\tdix{3}\ccc$&  & & &&&X&X&X\\ 
\hline
$\rm [S]_{tot}=3.0\tdix{-6}$&X  &X & X&X&X&&&\\ 
\hline
$\rm [S]_{tot}=1.3\tdix{-5}$&  & & &&&X&X&X\\ 
\hline
\end{tabular}
\label{tab:models}
\tablefoot{\tablefoottext{a} All these models are based on the
  \cite{legal2014a} model to which we have applied the modifications
  mentioned in the first column.}
\end{table*}

\subsection{Impact of a non-zero initial atomic hydrogen abundance}
\label{subsect:ini_chem_cond}

\begin{figure}
 \centering
\resizebox{\hsize}{!}{ \includegraphics{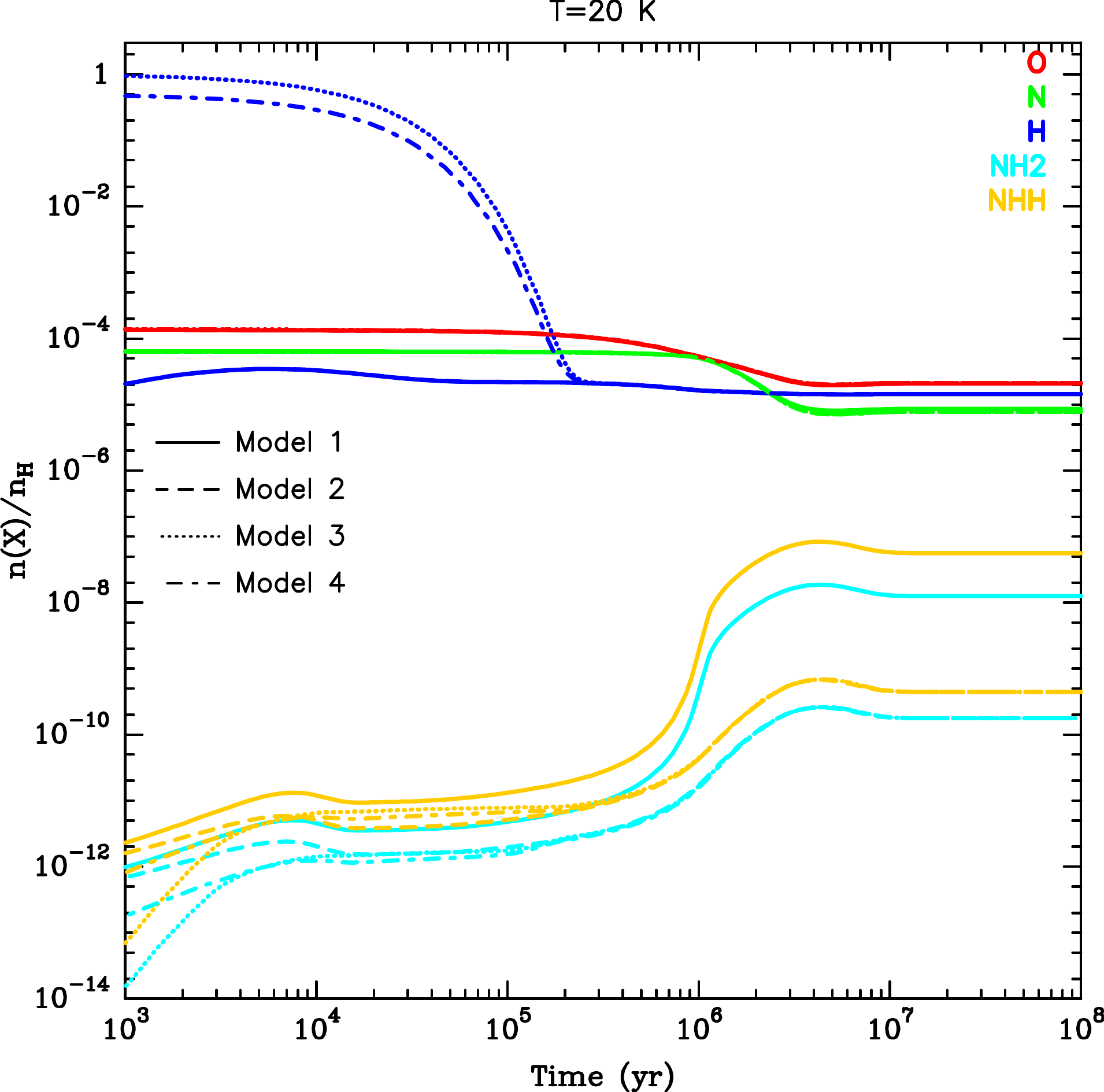}}
\caption{Evolution of the abundances of H, N, O, \ce{p-NH2} (NH2 in
  cyan), and \ce{o-NH2} (NHH in yellow) computed at 20~K as functions
  of time with Models 1, 2, 3, and 4. The different models are described
  in Table~\ref{tab:models}.}
 \label{fig:H_N_O_NH2_ab}
\end{figure}

\begin{figure}
  \centering
  \resizebox{\hsize}{!}{ \includegraphics{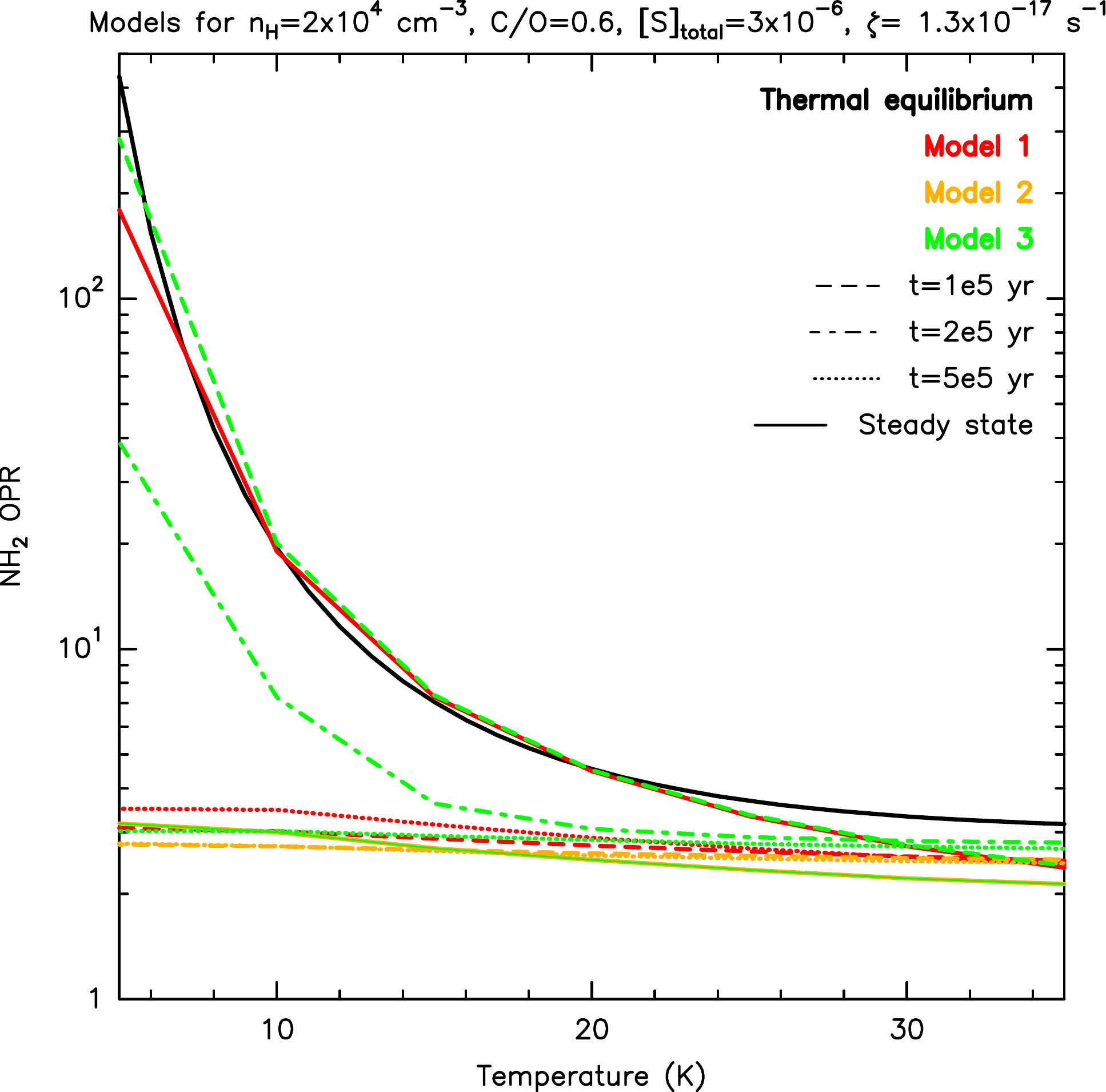}}
 \caption{ Same as Figure~\ref{fig:NH2-ROP_ture} with the addition of
   Model 3 and additional timescales.}
\label{fig:NH2-ROP_ture_3models}

\end{figure}

The extent of thermalization of the NH$_{2}$ OPR at low temperatures
will be increased by increasing the amount of atomic hydrogen in the
gas phase, using alterations commonly used in other models. One way to
do so is to run the model with the hydrogen elemental abundance
initially in its atomic form rather than fully molecular. Model 3
contains this change, and is otherwise identical with Model 2, as
described in Table~\ref{tab:models}. The amount of atomic hydrogen as
a function of time is increased up to a time of $\sim \dix{5}$~yr for
the model runs at 20~K, as displayed by the dotted lines in
Figure~\ref{fig:H_N_O_NH2_ab}, where an enhancement ranging up to
several orders of magnitude can be seen, such that the H abundance far
exceeds the abundances of O and N. As a consequence, the \ce{NH2}
destruction reactions by N and O are negligible compared with the
H-exchange reaction, which can then thermalize the \ce{NH2} OPR.
Figure~\ref{fig:NH2-ROP_ture_3models} shows the \ce{NH2} OPR as
function of temperature at thermal equilibirum and for Models 1, 2,
and 3 at different timescales. From this figure, one can see that the
low temperature thermalization of Model 1 at steady state is recovered
in Model 3 but only at earlier times. Above a timescale of $\sim
2\tdix{5}$~yr, the thermalization becomes less efficient due to the
decrease of the atomic hydrogen abundance.

\begin{figure}
 \centering
\resizebox{\hsize}{!}{ \includegraphics{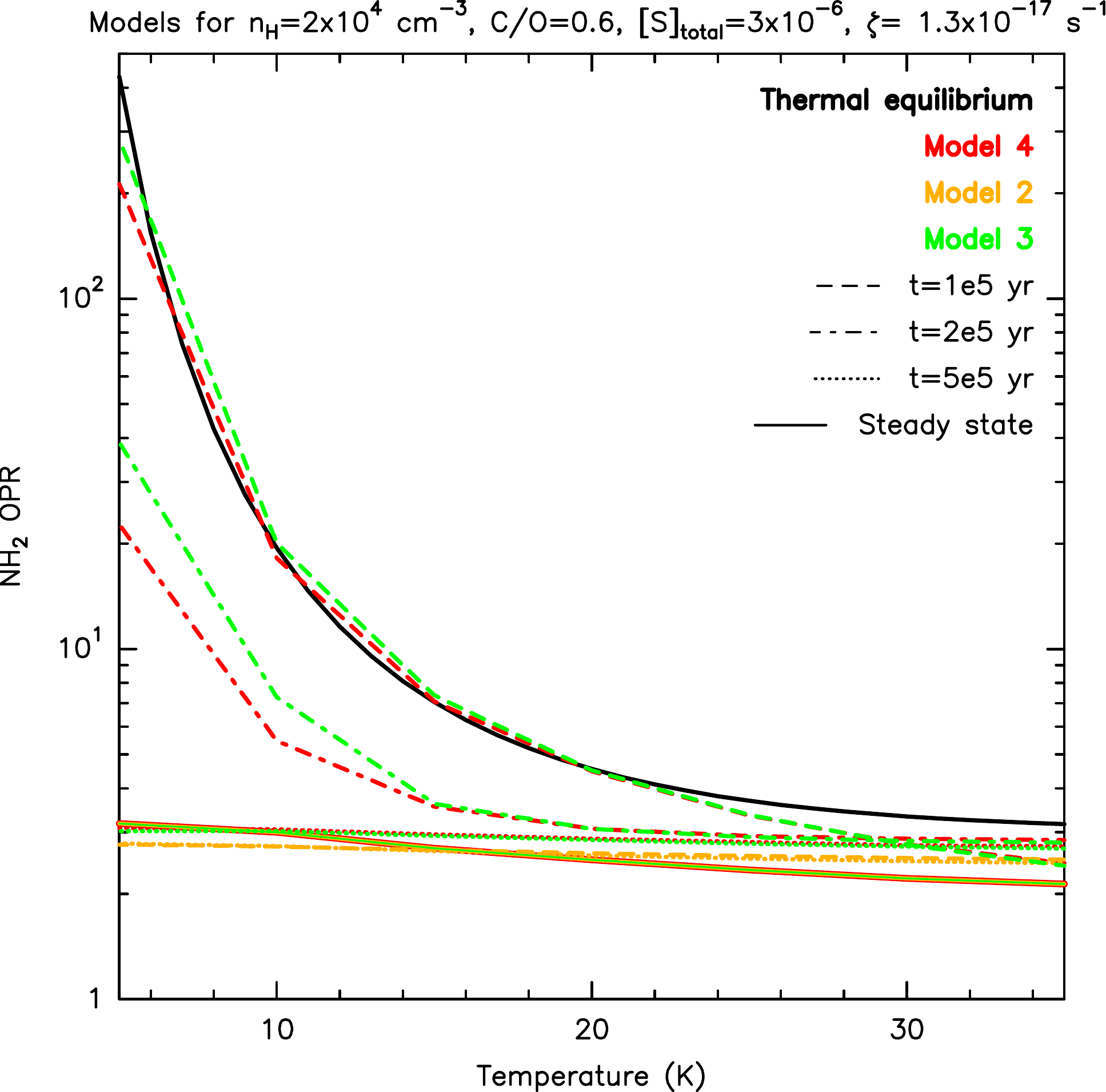}}
\caption{Same as Figure~\ref{fig:NH2-ROP_ture_3models} but with Model
  4 results replacing those of Model 1. This figure highlights the
  impact of the initial chemical condition on the OPR of NH$_2$ as a
  function of temperature.}
\label{fig:NH2-ROP_ture_3models_2}
\end{figure}

Since our astrochemical model is a pseudo-time dependent model,
meaning that the physical conditions are fixed for the entire
simulation as a function of time, it would probably be more realistic
to start with hydrogen initially half atomic and half molecular. This
constitutes Model 4, for which the OPR results are represented in
Figure~\ref{fig:NH2-ROP_ture_3models_2} and the H, N, O, o-NH$_{2}$
and p-NH$_{2}$ abundances at 20~K by the dashed-dotted lines in
Figure~\ref{fig:H_N_O_NH2_ab}. The results are quite similar to those
for Model 3, by comparison of the dotted and dashed-dotted lines in
the figure.

\subsection{Impact of the cosmic-ray ionization rate}
\label{subsect:Zeta}

\begin{figure}
 \centering
\resizebox{\hsize}{!}{ \includegraphics{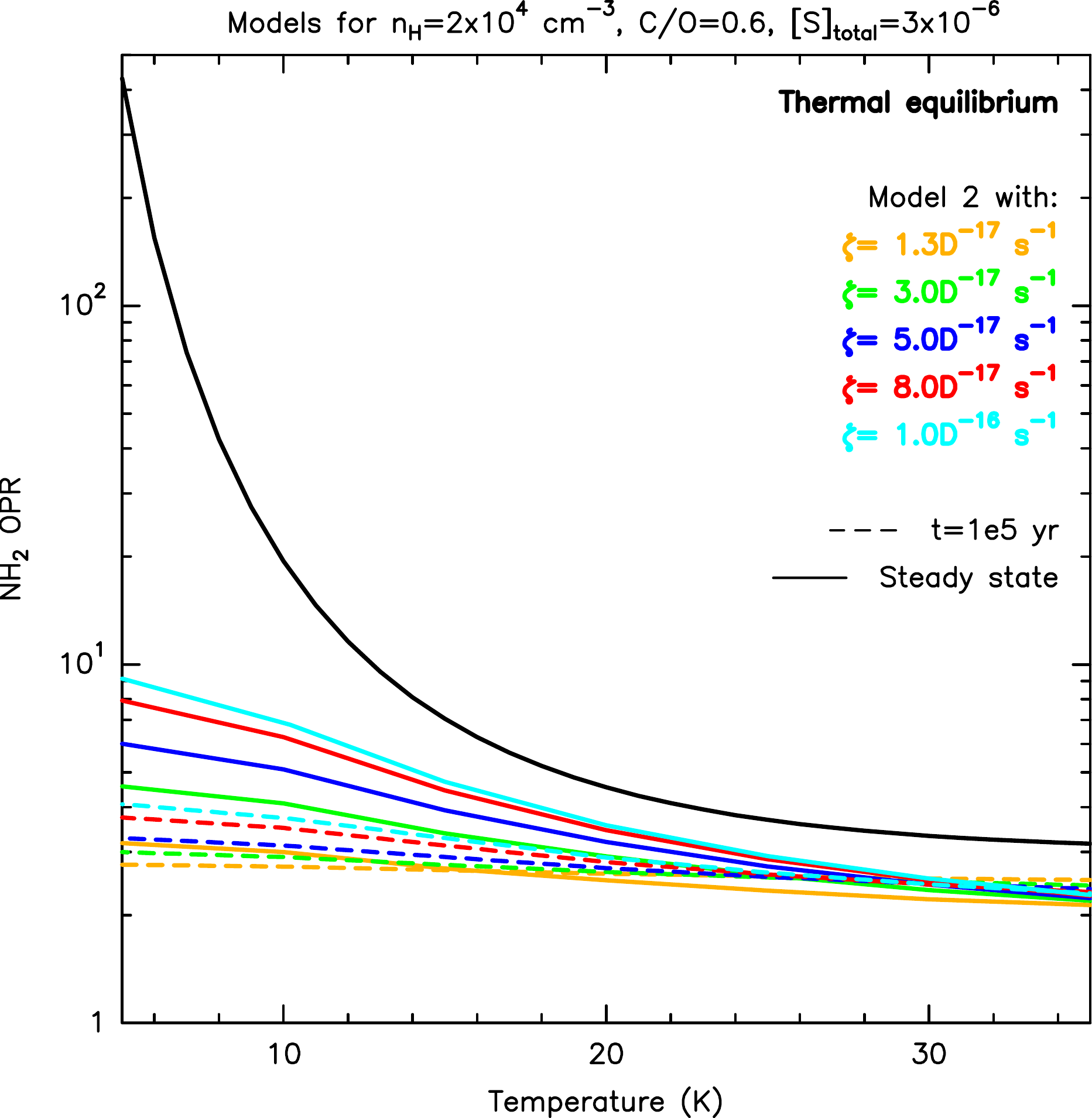}}
\caption{ NH$_2$ OPR computed with Model 2 for different cosmic-ray
  ionization rates.  The \ce{NH2}-OPR values are plotted at two
  different times: $1\times10^5$~yr and at steady state.}
 \label{fig:NH2-ROP_ture_zeta_impact}
\end{figure}

\begin{figure}
 \centering
\resizebox{\hsize}{!}{ \includegraphics{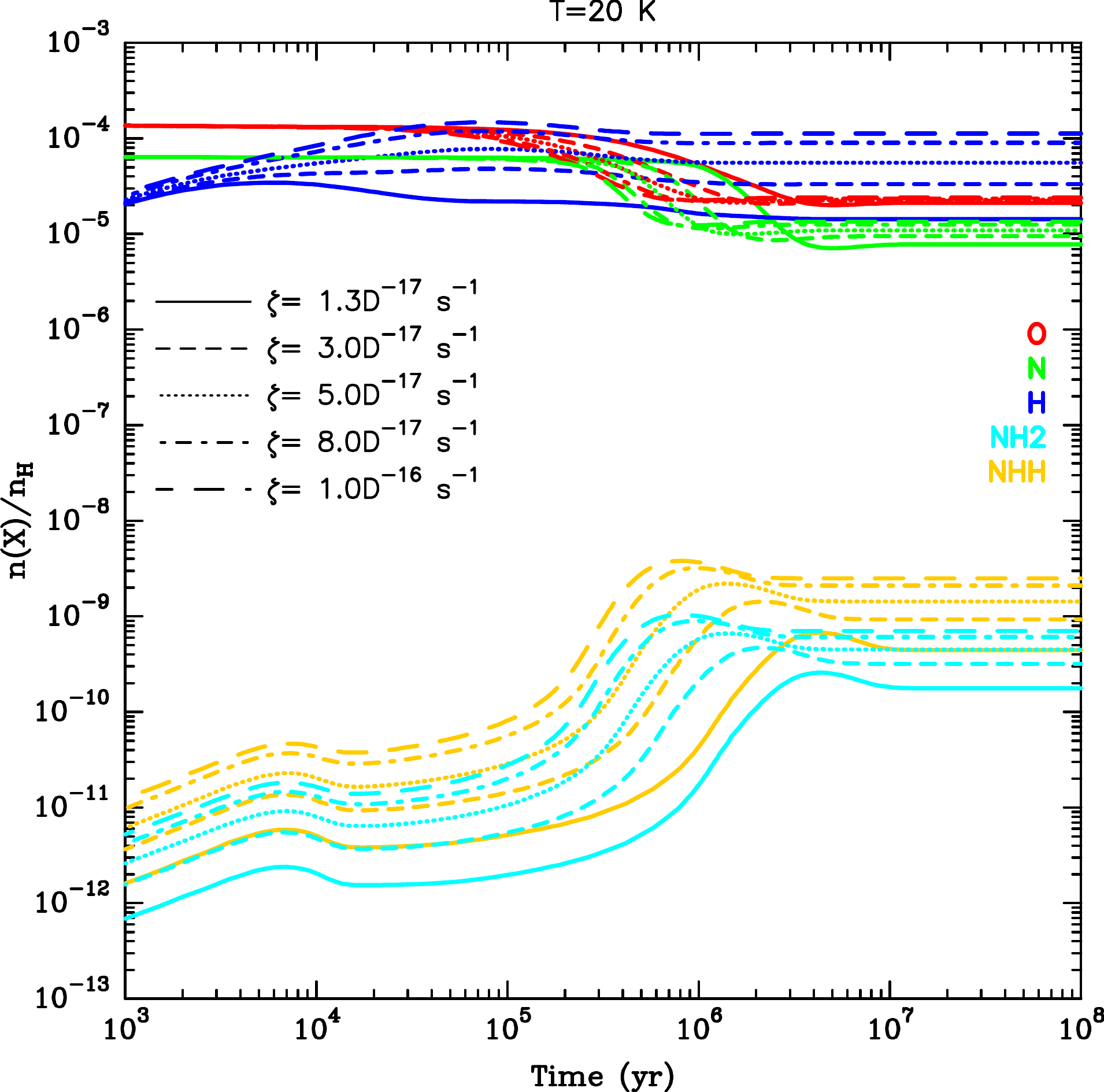}}
\caption{Evolution of the abundances of H, N, O , \ce{p-NH2} (NH2 in
  cyan) and \ce{o-NH2} (NHH in yellow) computed at 20~K with Model 2 as
  functions of time and for different cosmic-ray ionization
  rates. The details of Model 2 can be found in Table~\ref{tab:models}.}
 \label{fig:H_N_O_NH2_time_zeta_impact}
\end{figure}

Another possible way to increase the atomic hydrogen abundance in the
gas phase is to vary the cosmic-ray ionization rate, $\zeta$, which is
not well constrained in dense cold gas. We have varied $\zeta$ in
between the commonly used value of $1.3\tdix{-17}\pers$ and
$1\tdix{-16}\pers$, which lies at the upper limit for dark cores
\citep[see \eg][]{caselli1998}. The OPR values that we obtained by
running Model 2 with these different values of the cosmic-ray
ionization rate are displayed in
Figure~\ref{fig:NH2-ROP_ture_zeta_impact} as functions of temperature
at steady state and an earlier time. As can be seen, increasing the
cosmic-ray ionization rate increases the thermalization of the
\ce{NH2} OPR at low temperatures. The results in
Figure~\ref{fig:H_N_O_NH2_time_zeta_impact} show that increasing
$\zeta$ increases the atomic hydrogen abundance in the gas phase and
thus makes the \ce{H + NH2} H-exchange reaction more efficient
compared to the destruction reactions of \ce{NH2} by N and O.

Specifically, we find that increasing the cosmic-ray ionization rate
by one order of magnitude increases the abundance of hydrogen by
approximately the same amount, and also that even an increase to
$\zeta=3\tdix{-17}\pers$, a factor approximately two times the
standard value, allows the model to produce \ce{NH2}-OPR values above
3. This particular model is named Model 5 in
Tables~\ref{tab:models}. Interestingly, the increased thermalization
in Model 5 complements the increased thermalization produced by an
initial non-zero abundance of atomic hydrogen in that the impact of
the ionization rate starts to affect the hydrogen abundance at a few
$\dix{4}$~yr, as depicted in
Figure~\ref{fig:H_N_O_NH2_time_zeta_impact}, and so increases the
efficiency of the H-exchange reaction with time, contrary to the
effect observed when only increasing the initial atomic hydrogen
abundance.

\subsection{Translucent clouds}
\label{subsect:trans}

\begin{figure}
 \centering
\resizebox{\hsize}{!}{ \includegraphics{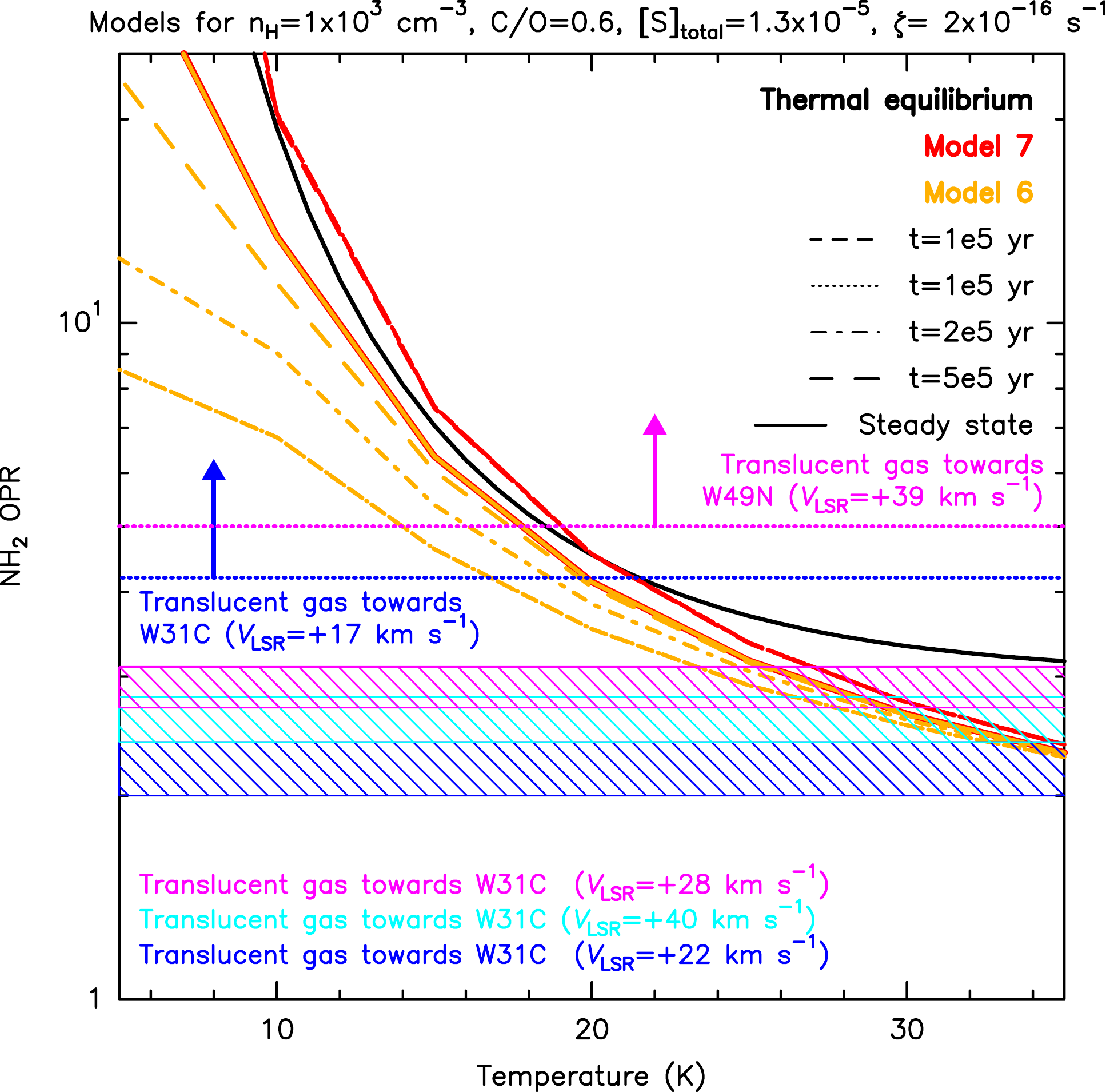}}
\caption{Calculated OPR of NH$_2$ computed as a function of
  temperature for a density of $n_\mathrm{H} = 1\times10^3$~cm$^{-3}$
  for translucent sources. Shown are OPR values for thermal
  equilibrium and Models 6-7 at assorted times. The hatched boxes
  represent the OPR measurements from \cite{persson2016} within their
  formal errors, for the translucent gas towards W31C found at three
  different gas velocities. The dotted horizontal lines with arrows
  mark the lower limits in the translucent gas towards W31C in blue
  and W49N in pink. The temperature range is limited to the 5~K - 35~K
  even though the observed ranges can be as large as 100~K.  See text.}
 \label{fig:NH2_ture_translucent_case}
\end{figure}

For the \ce{NH2} OPR observed in translucent gas towards W31C (from
2.0 to 3.1 and $\gtrsim 4.2$) and towards W49N ($\gtrsim 5$)
\citep{persson2016} we ran two more models, labeled Model 6 and Model
7. These models both contain a lower gas density of $ n_{\rm
  H}=1\tdix{3}\ccc$, a higher sulfur elemental abundance of $\rm
[S]_{tot}=1.3\tdix{-5}$, corresponding to the cosmic value
\citep{asplund2009}, and a higher $\zeta$ of $2 \times 10^{-16}$
s$^{-1}$. These values are more appropriate values for translucent to
diffuse gas. Model 6 has an initial abundance of hydrogen that is
totally molecular, while Model 7 starts with 50\% H and 50\%
H$_{2}$. The \ce{NH2}-OPR values obtained with these models as
functions of the temperature are shown in
Figure~\ref{fig:NH2_ture_translucent_case} for different timescales
from $\dix{4}$~yr to steady state, along with the observed values,
within their formal errors, represented by hatched boxes and by the
dotted pink and blue lines for the lower limits. Because the
constraints on the temperatures corresponding to the observations
other than OPR values are weak \citep{persson2016}, we choose in
Figure~\ref{fig:NH2_ture_translucent_case} to represent the observed
OPR values along a temperature range from 5 K to 35 K only, allowing
one to see how each model can fit the observations on over portions of
this range. From figures such as this, one can determine the
temperature ranges at a given timescale when the model OPR values are
in agreement with those of observed sources.

\section{Discussion}
\label{section:disc}
The temperature ranges for which each model reproduces the
NH$_{2}$-OPR values within their uncertainty ranges are tabulated in
Table~\ref{tab:opr_results}. The table is constructed in the following
format: the first column on the left lists the three types of sources
observed and studied: molecular envelopes, dense and cold cores, and
translucent gas.  In each category, we tabulate the observed NH$_{2}$
OPR and range of temperatures associated with the sources in the
regions W31C, W49N, W51, and G34. Unless the OPR values have very
large uncertainties, the observed temperature ranges likely pertain to
a diversity of regions not included in the OPR observations, which
have a tight constraint on the temperature range, corresponding to the
lowest values of the observed ranges. We then list the models that can
reproduce the observed OPR values over some temperature range, which
may or may not overlap with the observed range of temperatures.
Several timescales are listed for each model. For example, consider
the molecular envelope in W49N, which has an OPR of~3.5. This value
can be matched by Models 1, 4, and 5, but none of these models can
also match the high upper limit of the observed temperature ranges,
which is as high as 120~K. On the other hand, consider the cold core
in W51, where the OPR is measured to be 3.4. This can be matched by
Models 1, 4, and 5. The observed temperature range for W51 is
10-30~K. All three models present smaller temperature ranges within
this rather large observational range. Finally, we consider the
translucent gas in W49N. Here the OPR value of $\ge$ 5.0 is matched by
Models 1', 6, and 7. The observed temperature of $\le$ 15 K is matched
well by the three models. Note that Model 1' is similar to Model 1
except that the density, ionization rate and sulfur abundance are
those used for the translucent case, as described in
Table~\ref{tab:models}.

For those models with the thermalization of the OPR for NH$_{2}$
activated, we must also investigate how this activation changes the
degree of agreement with OPR values for sources at higher
temperatures, where these values are lower than three. As described in
Table~\ref{tab:opr_results}, Models 2, 4, and 5 for dense sources, and
Models 1', 6 and 7 for translucent sources generally fit the data,
despite the fact that the range of the observed temperature
uncertainties is large. Since the models presented in this study
always give smaller temperature ranges than the observed ones, and
towards the lower range of temperature, we are tempted to claim that
the OPR measurements do not pertain to those portions of the observed
clouds with temperatures high enough for the OPR to be only at the
statistical limit of three or below, which is roughly 40 K.

\begin{landscape}
\begin{table}[\!htb] 
\centering
\caption{Observed and calculated OPR values of NH$_{2}$ for assorted models with associated temperature ranges.}
\begin{tabular} {|c|c|c|cc|cccccccccc} 
  \hline\hline
  Source  & OPR\tablefoottext{a} & $T_\mathrm{K}$ &
  \multicolumn{2}{c|}{$T_\mathrm{range}$ for model 1\tablefoottext{b}} &\multicolumn{2}{c|}{$T_\mathrm{range}$ for model 2}&\multicolumn{4}{c|}{$T_\mathrm{range}$ for model 4} &\multicolumn{2}{c|}{$T_\mathrm{range}$ for model 5}\\
  &&&$5\times10^5$~yr &$t\gtrsim10^6$~yr&$(1-5)\times10^5$~yr&\multicolumn{1}{c|}{s.s.} &$1\times10^5$~yr &$2\times10^5$~yr&$5\times10^5$~yr&\multicolumn{1}{c|}{s.s.}&$1\times10^5$~yr&\multicolumn{1}{c|}{s.s.} \\
  & & (K)& (K) & (K) & (K) & \multicolumn{1}{c|}{(K)}& (K) & (K) & (K) & \multicolumn{1}{c|}{(K)}& (K) & \multicolumn{1}{c|}{(K)} \\
  \noalign{\smallskip}
  \hline
  \noalign{\smallskip}  
  Molecular envelopes \\ 
  \hline
  W31C & 2.5($\pm0.1$)
  &$30-50$\tablefoottext{c}&$28-35$&$32-35$&$\gtrsim$18&\multicolumn{1}{c|}{$17-23$}&--&--&--&\multicolumn{1}{c|}{$17-23$}&$\gtrsim22$&\multicolumn{1}{c|}{$25-29$}\\
  W49N &
  3.5($\pm0.1$)&$\sim$15-120\tablefoottext{d}&$5-12$&$24-25$&--&\multicolumn{1}{c|}{--}&$24-25$ & $15-17$&--&\multicolumn{1}{c|}{--}&--&\multicolumn{1}{c|}{$13-15$}\\   
  W51 & 2.7($\pm0.1$)&20-50\tablefoottext{e}&$23-28$&$29-32$&$\lesssim$18&\multicolumn{1}{c|}{$13-17$}&--&--&$\gtrsim$24&\multicolumn{1}{c|}{$13-17$}&$14-23$&\multicolumn{1}{c|}{$22-25$}  \\   
  G34 & 2.3($\pm0.1$)&20-70\tablefoottext{e,f}&$\gtrsim$35&$\gtrsim$35 &$\gtrsim$35&\multicolumn{1}{c|}{$23-32$} &--&--&--&\multicolumn{1}{c|}{$23-32$} &$\gtrsim$35&\multicolumn{1}{c|}{$29-35$}\\   
  \noalign{\smallskip}
  \hline
  \noalign{\smallskip} 
  Dense $\&$ cold core \\
  \hline
  \noalign{\smallskip} 
  W51 &   3.4($\pm0.1$) &10-30\tablefoottext{g}&$5-13$&$24-25$&--&\multicolumn{1}{c|}{--}&$24-26$ & $15-18$&--&\multicolumn{1}{c|}{--}&--&\multicolumn{1}{c|}{$14-16$}\\
  \noalign{\smallskip}
  \hline
  \hline
  \noalign{\smallskip} 
  Translucent gas & & & \multicolumn{2}{c|}{$T_\mathrm{range}$ for
    model 1'\tablefoottext{b}}&\multicolumn{4}{c|}{$T_\mathrm{range}$ for model
    6}&\multicolumn{2}{c}{$T_\mathrm{range}$ for model 7}
  &\multicolumn{2}{|c|}{\multirow{7}{*}{\diagbox[width=35mm,height=27.5mm]{}{}}}\\
  \hline
  \noalign{\smallskip}
  &&&$10^4$~yr &$\gtrsim 10^6$~yr &$10^4$~yr &$10^5$~yr &$2\tdix{5}$~yr
  &\multicolumn{1}{c|}{$\gtrsim 5\tdix{5}$~yr} &$10^4-5\tdix{5}$~yr&s.s. &\multicolumn{2}{|c|}{}\\
  \noalign{\smallskip}  
  \hline
  W31C & 2.2($\pm0.2$) & 30-100\tablefoottext{h} &$\gtrsim$27&$\gtrsim$34 &$\gtrsim$25&$\gtrsim$33 &$\gtrsim$34 &\multicolumn{1}{c|}{$\gtrsim$34} &$\gtrsim$35  &$\gtrsim$34 &\multicolumn{2}{|c|}{}\\
  & 2.9($\pm0.2$) & 20-100\tablefoottext{h} &$5-16$& $25-29$  &$5-11$& $23-28$& $25-29$&\multicolumn{1}{c|}{$26-30$}& $27-31$ &$26-30$ &\multicolumn{2}{|c|}{}\\
  & 2.6($\pm0.2$) & 25-75\tablefoottext{h} &$12-27$&$28-34$ &$5-25$& $27-32$& $28-33$& \multicolumn{1}{c|}{$29-34$}& $30-35$ &$29-34$&\multicolumn{2}{|c|}{}\\    
W31C & $\gtrsim$4.2 & 30-85\tablefoottext{h} &$\lesssim17$&$\lesssim21$&--&$\lesssim17$&$\lesssim19$&\multicolumn{1}{c|}{$\lesssim20$}&$\lesssim21$&$\lesssim20$&\multicolumn{2}{|c|}{}\\
W49N &  $\gtrsim$5.0 & $<$15\tablefoottext{h} &$\lesssim14$&$\lesssim19$&--&$\lesssim14$&$\lesssim16$&\multicolumn{1}{c|}{$\lesssim18$}&$\lesssim19$&$\lesssim18$&\multicolumn{2}{|c|}{}\\
 \hline 
\end{tabular}
\tablefoot{$T_\mathrm{K}$  and s.s. stand respectively for
  the observed temperatures and steady state.
  \tablefoottext{a} The tabulated errors are the formal errors from
  \cite{persson2016}. 
  \tablefoottext{b} The models 1 and 1' are similar to the models b
  for dense and translucent cases, respectively, described in \cite{persson2016}.
  \tablefoottext{c}{\citet{fazio1978} and \citet{mueller2002}.}
  \tablefoottext{d}{\citet{vastel2001}.}
  \tablefoottext{e}{\citet{vandertak2013}.}
  \tablefoottext{f}{Derived from NH$_3$ rotational transitions
    \citep{hajigholi2016}.}
  \tablefoottext{g}{Derived from CN and
    NH$_3$ rotational transitions \citep{mookerjea2014}.}
  \tablefoottext{h}{The  excitation temperature of the
    C\ion{I} 492~GHz line  \citep{gerin2015}.}
}
\label{tab:opr_results}
\end{table}
\end{landscape}

\section{Summary}

In this work we have reported quasi-classical trajectory calculations
for the \ce{H + NH2} H-exchange reaction showing that the reaction
occurs without a barrier and with a low-temperature rate coefficient
of approximately $1 \times 10^{-10}$ cm$^3$ s$^{-1}$. This result is
of importance because the \ce{H + NH2} H-exchange reaction was
introduced for the first time in astrochemical models in
\cite{persson2016} in order to allow the \ce{NH2} OPR to tend towards
the high thermal ratio at low temperatures as a function of time, with
the greatest effect at steady state. Therefore, this calculation
appears to confirm that suggestion. Indeed, as shown here, even lower
rate coefficients for the \ce{H + NH2} H-exchange reaction can still
raise the \ce{NH2} OPR so that it comes closer to the high thermal
ratio at low temperatures.

However, in this study, we included new rate coefficients and
processes for the destruction of \ce{NH2} by reactions with atomic
oxygen and nitrogen. This inclusion has decreased the calculated
\ce{NH2}-OPR value as a function of temperature and timescale to
values closer to those obtained before we included the \ce{H + NH2}
H-exchange reaction by destroying the NH$_{2}$ at a faster rate than
the exchange process. As a result, the NH$_{2}$-OPR calculations can
no longer reproduce those values obtained for sources where they are
observed to be greater than three. We have found several methods to
turn the thermalization of NH$_{2}$ back on, and so reproduce these
observed data. One possibility is to increase the amount of atomic
hydrogen in the gas phase. This can be done in a variety of ways. One
method is to change the initial abundance of hydrogen from purely
H$_{2}$ to purely atomic hydrogen, while a more physically reasonable
approach is to start with a mixture that is half atomic and half
molecular hydrogen. These methods are used in Models 3, 4 and 7, and
tend to be successful at early times before a few~$\dix{5}$~yr after
which the initial atomic hydrogen is mainly converted to its molecular
form. We also investigated the impact of varying the cosmic-ray
ionization rate, and found that increasing the cosmic-ray ionization
rate increases the abundance of atomic hydrogen, improves the \ce{H +
  NH2} reaction efficiency, and consequently the thermalization of the
\ce{NH2} OPR at low temperatures. Furthermore, the use of this second
method complements the first approach since the impact of the
ionization rate variation starts to affect the hydrogen abundance at a
few $\dix{4}$~yr, increasing the thermalization of the \ce{NH2} OPR
with increasing time. Models 1' and 5 to 7 contain cosmic ray
ionization rates higher than the standard value.

The two methods used here to increase the abundance of the atomic
hydrogen in the gas phase are of course not the only ways to do
so. For instance, we could decrease the efficiency of the formation of
\ce{H2} on grain surfaces, as suggested by \cite{cuppen2006}. Yet
another approach would be to increase the standard low-metal C/O
abundance ratio from 0.6 to higher values by lowering the oxygen
abundance, and so reducing the destruction rate of NH$_{2}$ by oxygen.

Because the temperatures of the sources studied here are not well
determined, the nature of the agreement between observations and our
calculations is the degree to which the ranges of temperatures
corresponding to the observed and calculated OPR values overlap at
physically reasonable timescales. In general, this agreement is
reasonable for observed OPR values both below and above three for
models that include some degree of NH$_{2}$-OPR thermalization, as can
be seen in Table~\ref{tab:opr_results}. Furthermore, our modeled
temperature ranges are smaller than the observed ones which could
indicate that measuring the \ce{NH2}-OPR in each environment could
help to constrain its temperature to smaller ranges towards the lower
values. But there is one major omission in our nearly pure gas-phase
treatment and that is the role of ortho-to-para conversion on granular
surfaces. As already mentioned in \cite{persson2016}, the efficiency
of this phenomenon will depend on the time of residence of each
species on the grain, on the shape of the grain surface, and on the
time of nuclear-spin conversion. However, the characteristic
nuclear-spin conversion times on grain surfaces are not yet well
constrained \citep{lebourlot2000,chehrouri2011,hama2013}. So it should
be very interesting to explore ortho-para conversion on grain
surfaces, as reported by \cite{bron2016} for the \ce{H2} OPR.

\begin{acknowledgements}
  We thank our anonymous referee for his/her comments that improved
  the manuscript.  R. L. and E. H. acknowledge the support of the
  National Science Foundation (US) for his astrochemistry program, and
  support from the NASA Exobiology and Evolutionary Biology program
  through a subcontract from Rensselaer Polytechnic
  Institute. C. X. and H. G. thank US Department of Energy for
  financial support (Grant No. DE-FG02-05ER15694). A. L. acknowledge
  partial support from the Scientific Foundation of Northwest
  University (Grant No. 338050068).
\end{acknowledgements}

\newpage

\bibliographystyle{aa}
\bibliography{H-exchange-withNH2-biblio}

\end{document}